\documentclass[usenatbib]{mnras}

\usepackage{newtxtext,newtxmath}
\usepackage{graphicx}
\usepackage{amsmath}
\usepackage{natbib}

\newcommand{\mb}{m_{\rm b}}

\newcommand{\ev}{\,{\rm eV}}

\newcommand{\cm}{\,{\rm cm}}

\newcommand{\kpc}{\,{\rm kpc}}
\newcommand{\Mpc}{\,{\rm Mpc}}

\newcommand{\s}{\,{\rm s}}

\newcommand{\yr}{\,{\rm yr}}
\newcommand{\Myr}{\,{\rm Myr}}
\newcommand{\Gyr}{\,{\rm Gyr}}
\newcommand{\erg}{\,\rm erg}
\newcommand{\ergs}{\,\rm erg\ s^{-1}}

\newcommand{\K}{\,{\rm K}}

\newcommand{\msun}{\,{\rm M_{\odot}}}
\newcommand{\zsun}{\,{\rm Z_{\odot}}}

\newcommand{\vc}{v_{\rm c}}

\newcommand{\TBD}[1]{\textbf{(TBD: #1)}}

\def \arcsec    {^{\prime\prime}}

\newcommand{\Mstar}{M_*}

\newcommand{\nH}{n_{\rm H}}
\newcommand{\nHavg}{{\overline n_{\rm H}}}
\newcommand{\nHcool}{\overline n_{\rm H;\ cool}}

\newcommand{\kms}{\,\rm km\ s^{-1}}

\newcommand{\Mvir}{M_{\rm vir}}

\newcommand{\vvir}{v_{\rm vir}}
\newcommand{\Rvir}{R_{\rm vir}}
\newcommand{\Tvir}{T_{\rm vir}}

\newcommand{\Rsonic}{R(\tcoolsh=\tff)}

\newcommand{\tcool}{t_{\rm cool}}

\newcommand{\Rhalf}{R_{1/2}}

\newcommand{\tff}{t_{\rm ff}}

\newcommand{\Mhalo}{\Mvir}

\newcommand{\Tc}{T^{({\rm s})}}

\newcommand{\rhocrit}{\rho_{\rm crit}}

\newcommand{\Deltac}{\Delta_{\rm c}}

\newcommand{\Omo}{\Omega_{\rm m,0}}

\newcommand{\rvir}{R_{\rm vir}}

\newcommand{\dex}{\,{\rm dex}}

\newcommand{\tcoolsh}{\tcool^{({\rm s})}}

\newcommand{\hi}{\ion{H}{I}}
\newcommand{\cii}{[\ion{C}{II}]}

\newcommand{\fg}{f_{\rm gas}}
\newcommand{\changed}[1]{{#1}}
\newcommand{\NHI}{N_\hi}

\newcommand{\NH}{N_{\rm H}}
\newcommand{\Rimp}{R_\perp}

\newcommand{\Lya}{\text{Ly$\alpha$}}

\newcommand{\fDLA}{{\rm CF}_{\rm DLA}}
\newcommand{\ZDLA}{Z_{\rm DLA}}
\newcommand{\vDLA}{v_{\rm 90;\ DLA}}
\newcommand{\NDLA}{N_{\rm DLA}}

\newcommand{\Rthick}{R_{\rm shielded}}

\newcommand{\RthickZME}{R_{\rm ZME02}}
\newcommand{\RDLA}{\Rimp(\fDLA=0.5)}

\newcommand{\nhss}{n_{\rm H; shielded}}

\defcitealias{Stern20a}{S20}
\defcitealias{Stern20b}{S21}
\defcitealias{ZhengMiraldaEscude02}{ZME02}
\title[Neutral CGM as high-redshift DLAs]{Neutral CGM as damped \Lya\ absorbers at high redshift}

\author[J. Stern et al.]{
\parbox{\textwidth}{
Jonathan Stern$^{1,2}$\thanks{E-mail: sternjon@tauex.tau.ac.il},
Amiel Sternberg$^{1,3,4}$, 
Claude-Andr{\'e} Faucher-Gigu{\`e}re$^{2}$,
Zachary Hafen$^{2,5}$,
Drummond Fielding$^{3}$,
Eliot Quataert$^{6}$,
Andrew Wetzel$^{7}$,
Daniel Angl{\'e}s-Alc{\'a}zar$^{8,3}$,
Kareem El-Badry$^{9}$,
Du\v{s}an Kere\v{s}$^{10}$
and Philip F. Hopkins$^{11}$
}
\vspace{0.4cm}\\
\parbox{\textwidth}{
$^1$School of Physics and Astronomy, Tel Aviv University, Tel Aviv 69978, Israel\\
$^2$Department of Physics \& Astronomy and CIERA, Northwestern University, 1800 Sherman Ave, Evanston, IL 60201, USA \\
$^3$Center for Computational Astrophysics, Flatiron Institute, 162 5th Ave, New York, NY 10010, USA \\
$^4$Max-Planck-Institut für extraterrestrische Physik (MPE), Giessenbachstr., D-85748 Garching, FRG \\
$^5$Center for Cosmology, Department of Physics and Astronomy, 4129 Reines Hall, University of California Irvine, CA 92697, USA \\
$^6$Department of Astrophysical Sciences, Princeton University, Princeton, NJ 08544, USA \\
$^7$ Department of Physics \& Astronomy, University of California, Davis, CA 95616, USA \\
$^8$Department of Physics, University of Connecticut, 196 Auditorium Road, U-3046, Storrs, CT 06269-3046, USA \\
$^9$Astronomy Department and Theoretical Astrophysics Center, University of California Berkeley, Berkeley, CA 94720, USA \\
$^{10}$Department of Physics and Center for Astrophysics and Space Science, University of California at San Diego, 9500 Gilman Drive, La Jolla, CA 92093, USA\\
$^{11}$TAPIR, Mailcode 350-17, California Institute of Technology, Pasadena, CA 91125, USA
}}
\vspace{-40pt}

\date{\vspace{-20pt}Accepted XXX. Received YYY; in original form ZZZ}

\pubyear{2020}

\begin{document}
\label{firstpage}
\pagerange{\pageref{firstpage}--\pageref{lastpage}}
\maketitle

\begin{abstract}
Recent searches for the hosts of $z\sim4$ damped \Lya\ absorbers (DLAs) have detected bright galaxies at distances of tens of kpc from the DLA. 
Using the FIRE-2 cosmological zoom simulations, we argue that these relatively large distances are due to a predominantly cool and neutral inner circumgalactic medium (CGM) surrounding high-redshift galaxies. 
The inner CGM is cool because of the short cooling time of hot gas in $\lesssim10^{12}\msun$ halos, which implies that accretion and feedback energy are radiated quickly, while it is neutral due to high volume densities and column densities at high redshift which shield cool gas from photoionization. 
Our analysis predicts large DLA covering factors ($\gtrsim50\%$) out to impact parameters $\sim 0.3\left(\left(1+z\right)/5\right)^{3/2}\Rvir$ from the central galaxies at $z\gtrsim1$, 
equivalent to a proper distance of $\sim 21\,M_{12}^{1/3} \left(\left(1+z\right)/5\right)^{1/2}\kpc$ ($\Rvir$ and $M_{12}$ are the halo virial radius and mass in units of $10^{12}\msun$, respectively). 
This implies that DLA covering factors at $z\sim4$ may be comparable to unity out to a distance $\sim 10$ times larger than stellar half-mass radii. 
A predominantly neutral inner CGM \changed{in the early universe} suggests that its mass and metallicity can be directly constrained by absorption surveys, without resorting to the large ionization corrections as required for ionized CGM. 
\end{abstract} 


\section{Introduction}
Observations of damped \Lya\ absorbers along lines of sight to background quasars (DLAs, with \hi\ columns $\NHI > 2\cdot10^{20}\cm^{-2}$) 
provide some of the most stringent observational constraints on galaxy formation physics in the early post-reionization Universe. Statistical DLA samples are now available out to redshifts $z\sim5$ \citep[e.g.][]{ProchaskaWolfe09,Noterdaeme12,Ho20}, revealing the distribution of \hi-rich structures in and around galaxies far fainter than possible with emission surveys \citep{Wolfe05}. 
Metal line absorption features from DLAs constrain their metallicity and kinematics, and thus the DLA enrichment histories and the potential wells in which DLAs reside \citep{ProchaskaWolfe97, KulkarniFall02, Ledoux06, Pontzen08, Rafelski12,Rafelski14}. 
However, the origin of high-$z$ DLAs is still uncertain.
The DLAs may originate in central \hi-discs \citep{ProchaskaWolfe97, Fynbo08, Berry14}, in `protogalactic clumps' \citep{Haehnelt98}, 
in cosmic filaments \citep{FaucherGiguere11b,Fumagalli11}, in satellites and neutral outflows \citep{Rhodin19}, or in quasi-spherical cosmological inflows \citep{Theuns21}.

A new window on high-$z$ DLAs has recently been opened by the Atacama Large Millimeter/Submillimeter Array (ALMA), which can potentially detect \cii\ $158\mu{\rm m}$ fine-structure emission from associated galaxies.  \cite{Neeleman17,Neeleman19} detected \cii\ at angular separations of $2-4\arcsec$ from five high-metallicity DLAs at $z\sim4$, corresponding to proper distances of $20-40\kpc$. While this is comparable to typical DLA -- galaxy distances at $z\lesssim1$ \citep{ChenLanzetta03,Rahmani16,MollerChristensen20}, at $z\sim4$ this size scale is a substantial fraction of the halo virial radius ($\Rvir\approx60\kpc$ for a $\sim10^{12}\msun$ halo). This suggests that high-$z$ DLAs originate either in the CGM or in as yet undetected satellites galaxies. 

In this paper we use the FIRE-2 cosmological zoom simulations \citep{Hopkins18} to examine the contribution of the volume-filling phase of the circumgalactic medium (CGM) to the DLA population, where by `volume-filling phase' we refer to gas which permeates the halo volume rather than gas associated with subhalos or narrow filaments. We show that this phase is predominantly neutral with a large DLA covering factor at radii where two conditions are satisfied: 
(1) the cooling time of virial temperature gas $\tcoolsh$ is shorter than the free-fall time $\tff$, implying that heat from accretion and feedback is quickly radiated away, and 
(2) densities are sufficiently high that cool gas is shielded from ambient ionizing radiation. 
Our analysis builds on \citet[hereafter \citetalias{Stern20b}]{Stern20b} where we studied condition (1) in the FIRE simulations. We found that in $\lesssim10^{12}\msun$ halos in FIRE the inner CGM volume has $\tcoolsh<\tff$ and a volume-weighted temperature well below virial. In this paper we discuss the implications of this theoretical result for DLA observations. 
Our study also supplements previous FIRE results on CGM absorbers with lower \hi\ columns \citep{FaucherGiguere15,FaucherGiguere16,Hafen17}.

Our paper is organized as follows. 
In section~\ref{s:fire} we briefly summarize the relevant aspects of the FIRE simulations. 
In section~\ref{s:analytic} we explore the two conditions for a neutral CGM in FIRE, and demonstrate in section~\ref{s:CF} that these conditions are associated with an order-unity DLA covering factor.
Section~\ref{s:obs} compares the predictions of the FIRE simulations with observations of high-redshift DLAs. We discuss our results in section~\ref{s:discussion} and summarize in section~\ref{s:summary}. We assume a flat $\Lambda$CDM cosmology with Hubble constant $H_0=67\kms\Mpc^{-1}$, $\Omo=0.32$, and baryon mass fraction $\Omega_{\rm m}/\Omega_{\rm b}=0.158$ (\citealt{planck18}).

\section{FIRE-2 zoom simulations}\label{s:fire}

The Feedback In Realistic Environments \citep[FIRE,][]{Hopkins14,Hopkins18} project\footnote{\url{https://fire.northwestern.edu/}} was developed to explore the role of feedback in cosmological simulations of galaxy formation. We utilize the second version of these simulations, FIRE-2, which uses the multi-method gravity and hydrodynamics code GIZMO\footnote{\url{http://www.tapir.caltech.edu/\~phopkins/Site/GIZMO.html}} (\citealt{Hopkins15}) in its meshless finite-mass mode (MFM). Gravity is solved using a modified version of the Tree-PM solver similar to GADGET-3 \citep{Springel05}. 
Heating and cooling processes include metal line cooling, free-free emission, photoionization and recombination, Compton scattering with the cosmic microwave background, collisional and photoelectric heating by dust grains, and molecular and fine-structure cooling at low temperatures ($10-10^4\K$). 
Star formation occurs in self-gravitating, self-shielded molecular gas with $\nH>1000\cm^{-3}$, while the sub-grid implementation of feedback processes from stars includes radiation pressure, heating by photoionization and photoelectric processes, and energy, momentum, mass, and metal deposition from supernovae and stellar winds. AGN feedback is not included. A full description of the FIRE-2 simulations appears in \cite{Hopkins18}.

 \begin{figure}
\includegraphics{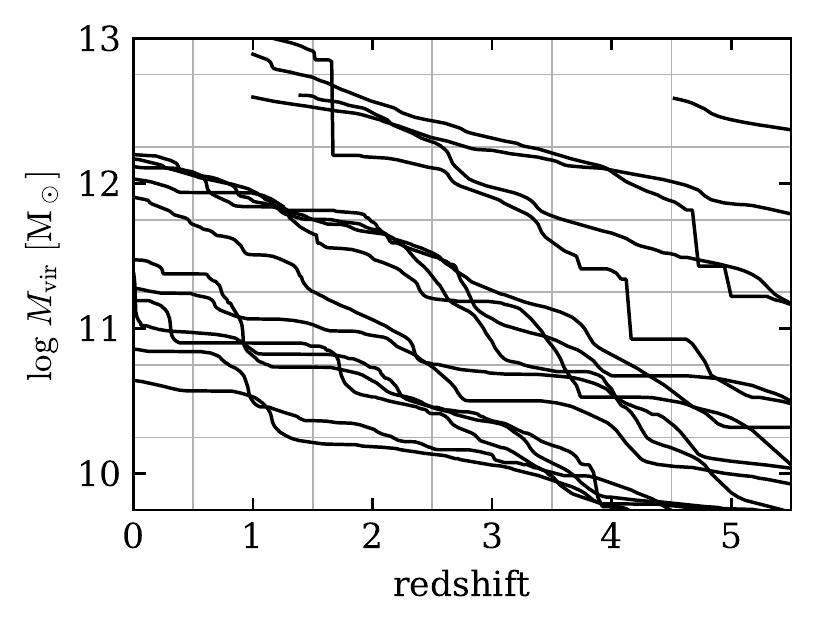}
\caption{
Tracks of central halo mass versus redshift in the 16 FIRE zoom simulations analyzed in this work. Grid lines delineate bins in $\Mhalo$ and $z$ used below. 
}
\label{f:mass tracks}
 \end{figure}

In the analysis we project \hi\ volume densities calculated in FIRE to predict observed \hi\ columns along sightlines to background sources. The \hi\ fraction of resolution element $a$ is derived in FIRE by assuming equilibrium between recombination, collisional ionization, and photoionization. Photoionization is by a spatially-uniform UV background (UVB) from \citet{FaucherGiguere09} and local hot stars. The contribution of a local stellar particle $b$ to the ionizing flux at resolution element $a$ is derived from the STARBURST99 stellar population models \citep{Leitherer99}, after accounting for absorption by resolution elements neighboring $b$ \citep[see appendix E in][]{Hopkins18}
and an approximation for absorption near $a$ as described next. The effect of other intervening absorbers between $b$ and $a$ on the stellar flux are neglected.
At resolution element $a$ the incident UVB  + stellar radiation is attenuated by a factor $e^{-\nH / \nhss}$, where the shielding density $\nhss$ is calculated from
\begin{equation}\label{e:nhss}
  \nhss = 0.0123\, \Gamma_{-12}^{2/3.} T_4^{0.173} \cm^{-3} ~,
\end{equation}
where $\Gamma\equiv 10^{-12}\Gamma_{-12} \s^{-1}$ is the photoionization rate and $T_4\equiv T/10^4\K$. 
This density threshold for shielding is derived from the assumption that the size of the structure which includes resolution element $a$ is equal to its Jeans length \citep[see appendix \ref{a:Jeans}]{Schaye01}. Using this approximation to attenuate the incident flux has been shown to reproduce the neutral columns derived from post-processing the simulations with full radiative transfer \citep[see][and further discussion in section~\ref{s:discussion}]{FaucherGiguere10, FaucherGiguere15, Rahmati13}. 

\changed{
We analyze the 16 FIRE simulations discussed in \citetalias{Stern20b}, chosen so their central halos span a diverse range in $(\Mhalo,z)$-space, as shown in Figure~\ref{f:mass tracks}. These include five massive halos from \cite{anglesalcazar17b} and \cite{ma18} for which $\Mhalo$ exceeds $10^{12}\msun$ at $z>2$, and eleven halos with $\Mvir(z=0)$ between $\sim10^{10.5}$ to $\sim10^{12}\msun$ \citep[from][]{Wetzel16,GarrisonKimmel17,GarrisonKimmel18,Hopkins18,Chan18,elbadry18a}. 
The mass of a baryonic resolution element is $\mb=33\,000-57\,000\msun$ in the five massive halos and $\mb=2100-7100\msun$ in the lower mass galaxies. We refer the reader to table~1 in \citetalias{Stern20b} and the original references for additional details on individual simulations. When binning is required below, we divide the simulation snapshots (separated by $25-30\Myr$) into bins of $\Delta\Mvir=0.5\dex$ and $\Delta z=1$ as shown in Figure~\ref{f:mass tracks}. 
}

Using the Amiga Halo Finder (\citealt{Knollmann09}) we identify $\Mhalo$ and $\Rvir$ of the main halo in each simulation snapshot based on the virial overdensity definition of \cite{bryan98}.  The virial temperature $\Tvir$ is then calculated from
\begin{equation}\label{e:Tvir}
 \Tvir \equiv \frac{\mu m_{\rm p} G \Mhalo}{2k_{\rm B} \Rvir} \approx 2.6\cdot10^6 M_{12}^{2/3}\left(\frac{1+z}{5}\right)^{0.9} \K
\end{equation}
where $\mu\approx0.62$ is the mean molecular weight, $m_{\rm p}$ is the proton mass, $G$ is the gravitational constant, $k_{\rm B}$ is the Boltzmann constant, and $M_{12}\equiv\Mhalo/10^{12}\msun$. The numerical estimate of $\Tvir$ is based on the approximation of $\Rvir$ given by eqn.~(\ref{e:rvir}) in the appendix.

\section{Conditions for a neutral CGM}\label{s:analytic}

Two conditions must be met for the CGM to have a large neutral fraction and thus also a large DLA covering factor. 
First, gas shocked to a temperature $\sim\Tvir$ must cool rapidly, otherwise the volume-filling phase would be hot and collisionally ionized. 
Second, any cool CGM needs to be shielded from the UVB and local stars, otherwise it would be photoionized. In this section we explore both conditions and demonstrate how they occur in the FIRE simulations.

\subsection{Cool gas filling fraction in the CGM}

\newcommand{\fv}{f_{\rm cool}}
\newcommand{\fcool}{f_{\rm cool}}
\newcommand{\frho}{f_{\rm cool;\ m}}

\begin{figure*}
\includegraphics{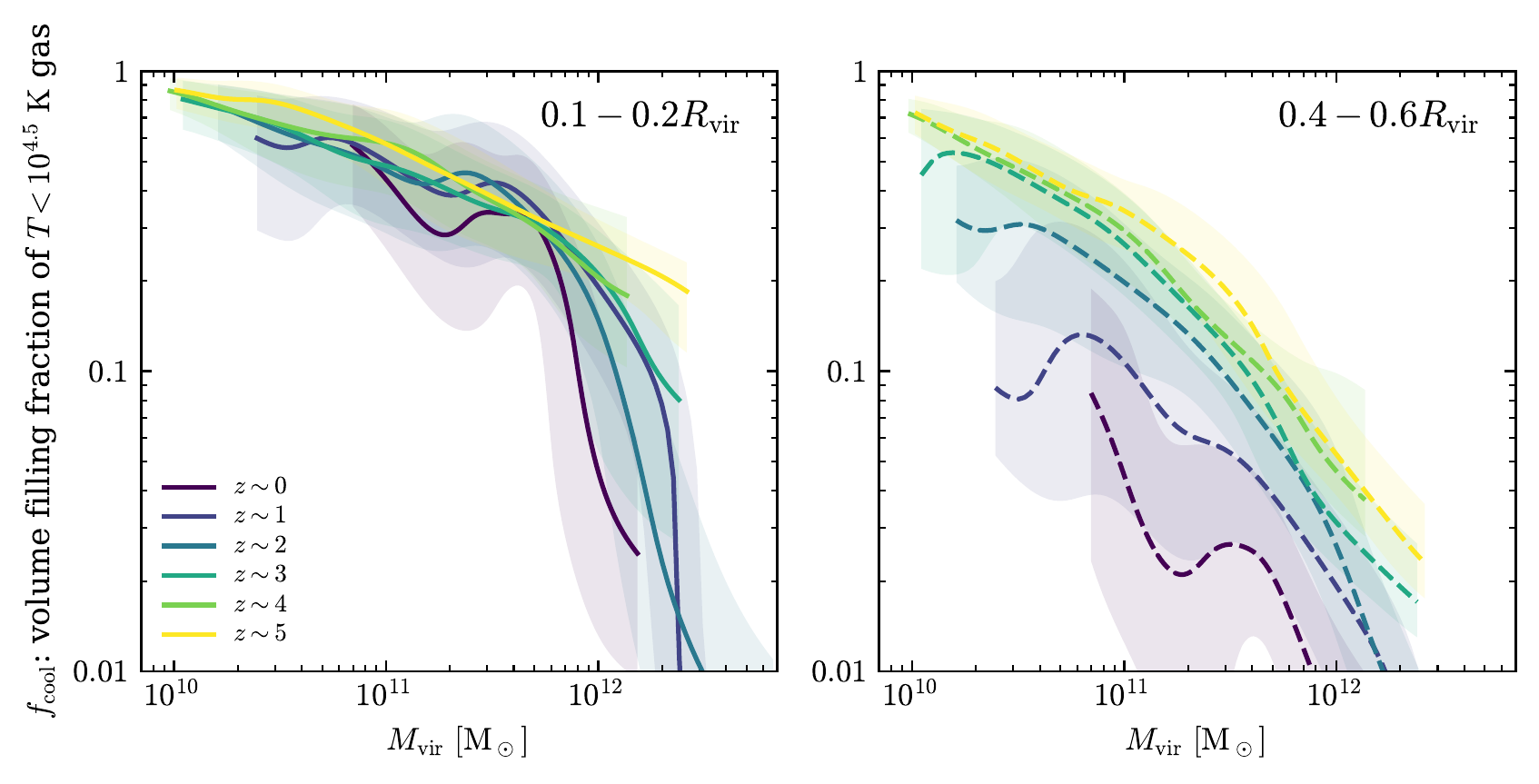}
\caption{The volume filling fraction of cool gas versus halo mass in FIRE, at inner CGM radii ($0.1<r/\Rvir<0.2$, \emph{left}) and at outer CGM radii ($0.4 <r/\Rvir<0.6$, \emph{right}). Colored lines and bands denote the median relation and $\pm1\sigma$ range for all snapshots within a redshift bin centered on the value noted in the legend. 
\changed{
At all redshifts the inner CGM typically has $\fv\gtrsim0.3$ in halos with mass $<5\cdot10^{11}\msun$, while $\fv$ tends to drop for halos with mass above $10^{12}\msun$. In the outer CGM large cool gas filling fractions of $\fv\gtrsim0.3$ are apparent in $<10^{11}\msun$ halos at $z>2$.
}}
\label{f:fV cool Mhalo}
\end{figure*}

\begin{figure*}
\includegraphics{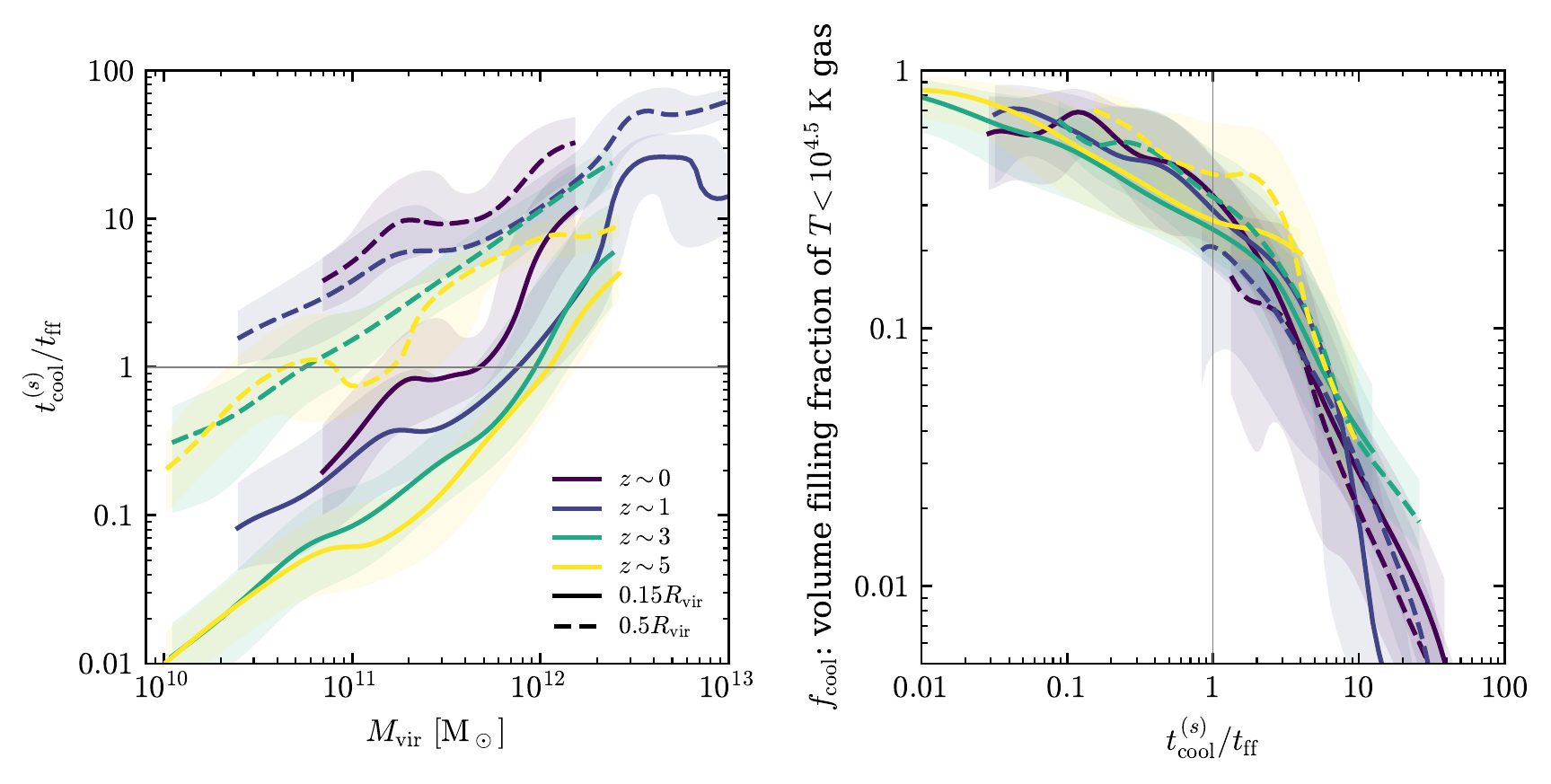}
\caption{
\textbf{Left:} The relation between halo mass and the ratio of the hot gas cooling time to free-fall time in FIRE. Solid and dashed lines correspond to measurements of $\tcoolsh/\tff$ at inner and outer CGM radii, respectively. 
Colored lines and bands plot the median and dispersion $(\pm1\sigma)$ for each redshift bin, as in Fig.~\ref{f:fV cool Mhalo}. Note that $\tcoolsh/\tff$ increases with halo mass and with halo radius, while it mildly decreases with increasing redshift. 
\textbf{Right:} Volume filling fraction of cool gas versus $\tcoolsh/\tff$. The ratio $\tcoolsh/\tff$ is estimated at the same radii as in the left panel, while $\fv$ is calculated within a shell spanning either $0.1<r/\Rvir<0.2$ (solid) or  $0.4<r/\Rvir<0.6$ (dashed). 
\changed{
The value of $\fv$ drops from $\approx0.6$ at $\tcoolsh\approx0.1\tff$ to $\fv\lesssim0.3$ at $\tcoolsh\approx10\tff$.
} 
This relation is rather independent of redshift and radius, suggesting that $\tcoolsh/\tff$ is the dominant parameter determining $\fv$. 
}
\label{f:fV cool}
\end{figure*}

\begin{figure}
\includegraphics{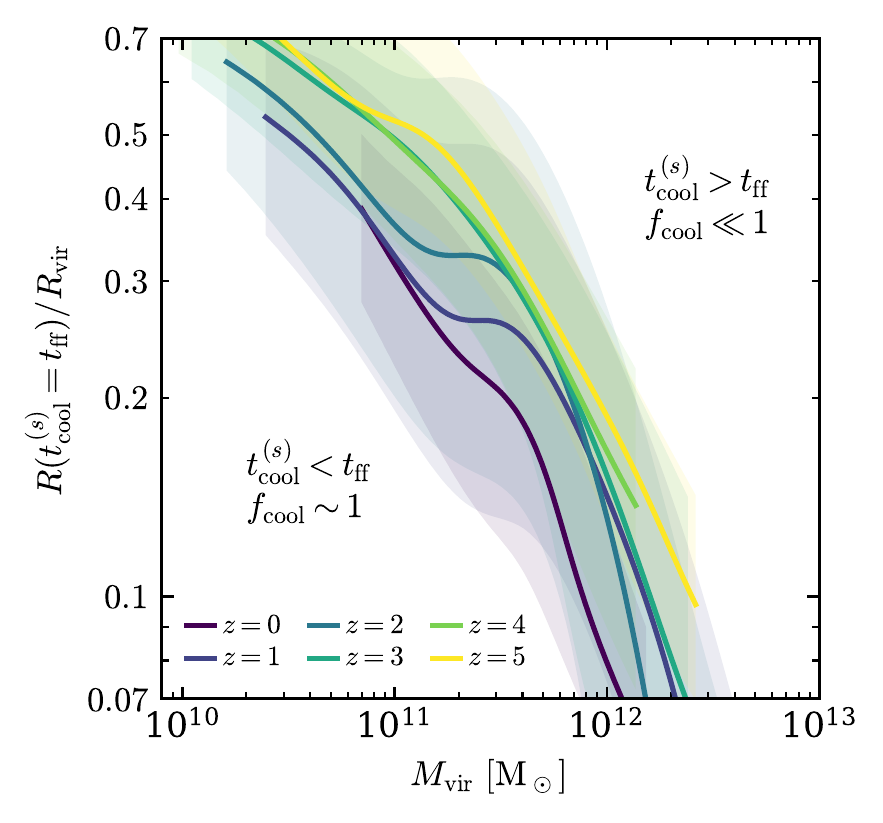}
\caption{The radius where $\tcoolsh=\tff$ versus halo mass and redshift. Colored lines and bands denote the median relation and $\pm1\sigma$ range for each redshift bin, as in Fig.~\ref{f:fV cool Mhalo}. The value of $R(\tcoolsh=\tff)$ is beyond $0.5\Rvir$ in halos with mass $\lesssim10^{11}\msun$ and drops to disc scales at $\Mhalo\gtrsim10^{12}\msun$. 
}
\label{f:Rsonic}
\end{figure}

The first condition for a neutral CGM, that hot gas cools rapidly, is expected when the cooling time of the hot phase $\tcoolsh$ is shorter than the free-fall time $\tff$. This has been demonstrated in idealized hydrodynamic simulations, in which the CGM becomes predominantly cool and supported by turbulence / bulk motions when $\tcoolsh\ll\tff$ \citep{McCourt12,Fielding17, Stern20a, Lochhaas20}, even when feedback heating is included. Similarly, in \citetalias{Stern20b} we showed that at radii where $\tcoolsh\lesssim\tff$ in the FIRE cosmological simulations the volume-weighted CGM temperature is $\ll\Tvir$. Here we further explore this result by calculating the volume filling fraction of cool gas $\fv$  as a function of $\tcoolsh/\tff$ and other halo parameters in FIRE. We define gas as `cool' if its temperature is below $10^{4.5}\K$, and limit the analysis in this section to snapshots with $\Tvir > 10^5\K$ (i.e., $\Mhalo>8\cdot10^9\msun$ at $z=4$) so the distinction between cool gas and virial temperature gas is well-defined. The chosen threshold of $T<10^{4.5}\K$ includes gas which is heated by photoionization, and also allows for some further heating by compression. As we discuss below, such gas would be neutral if shielded from photoionizing radiation. We emphasize that $\fcool$ is defined as a volume-fraction rather than a mass fraction, in order to focus on the volume-filling CGM phase while minimizing the effects of gas associated with narrow filaments and satellites.

We begin by exploring how $\fv$ depends on $\Mhalo$, $z$, and CGM radius $r$. 
To calculate $\fv(r)$ in a simulation snapshot we sum the volume 
of all resolution elements with $T<10^{4.5}\K$ that are within a given radial shell, and divide by the shell volume. 
Figure~\ref{f:fV cool Mhalo} plots $\fv$ as a function of halo mass for shells that span $0.1<r/\Rvir<0.2$ (left panel) and $0.4<r/\Rvir<0.6$ (right panel). Colored lines and bands plot the medians and dispersions $(\pm1\sigma)$ for all snapshots in a given redshift bin, where bins span $z_i-0.5<z<z_i+0.5$ for $z_i$ in $\{0,1,2,3,4,5\}$ (see Fig.~\ref{f:mass tracks}). 
Medians and dispersions are estimated by applying a Gaussian kernel density estimator \citep{Scott92} on measurements of $(\fv(r), \Mhalo)$ in all snapshots in the bin, using a kernel bandwidth of $0.3\dex$. The dispersion thus includes both variability for a given halo and halo-to-halo variation, though note the latter may be underestimated due to our relatively small halo sample size. 
The left panel shows that in the inner CGM $\fv$ is a strong function of $\Mhalo$, decreasing from near unity at $\Mhalo\sim10^{10}\msun$ to near zero at $\Mhalo\gtrsim10^{12}\msun$, with no strong dependence on redshift. 
At larger CGM radii shown in the right panel $\fv$ depends both on halo mass and on redshift, with $\fv>30\%$ apparent mainly in $\lesssim10^{11}\msun$ halos at $z\gtrsim2$.

Next, to compare $\fv$ with $\tcoolsh/\tff$ we estimate $\tff$ via 
\citep[e.g.,][]{McCourt12}
\begin{equation}\label{e:tff}
 \tff=\frac{\sqrt{2}r}{\vc} ~,
\end{equation}
where $\vc=\sqrt{GM(<r)/r}$ is the circular velocity and $M(<r)$ is the total mass enclosed within radius $r$. The estimate of the cooling time of shocked gas $\tcoolsh$ follows \citetalias{Stern20b} (see eqn.~7 there)
\begin{equation}\label{e:tcool}
 \tcoolsh\equiv\tcool(\Tc,P_{\rm HSE},Z,z)~,
\end{equation}
where $\tcool$ is the cooling time based on the \cite{Wiersma09} tables, 
$\Tc\equiv0.6\mu m_{\rm p}\vc^2(r)/k_{\rm B}$ is comparable to the virial temperature, 
$P_{\rm HSE}$ is the thermal pressure assuming hydrostatic equilibrium conditions, and $Z$ is the shell-averaged metallicity in the snapshot. To estimate $P_{\rm HSE}$ we assume it equals the spherically-averaged weight of overlying gas in the snapshot (eqn.~12 in \citetalias{Stern20b}). 
Equation~(\ref{e:tcool}) thus approximates the cooling time in a hot, thermal pressure-supported CGM, regardless of whether such a `virialized' CGM actually exists in the snapshot.

We checked that estimating $\tcoolsh$ based on the shell-averaged gas density rather than based on $P_{\rm HSE}$ yields similar results. This follows since when thermal pressure is absent, bulk motions and turbulence provide support against gravity (see below) and the resulting densities which go into the $\tcoolsh$ calculation are comparable. 
We refer the reader to \citetalias{Stern20b} for a discussion of $\tcoolsh$ and its comparison to other estimates of the cooling time in the simulations. 


The left panel of Figure~\ref{f:fV cool} compares $\tcoolsh/\tff$ with $\Mhalo$, using solid and dashed lines for the inner and outer CGM, respectively, and different colors for different redshifts as in Fig.~\ref{f:fV cool Mhalo}. The ratio $\tcoolsh/\tff$ increases strongly with $\Mhalo$ and decreases relatively weakly with increasing redshift. The strong dependence on halo mass is mainly a result of the increase of $\Tvir$ with $\Mhalo$, while the lack of a strong dependence on redshift is due to the decrease in gas density with time roughly cancelling the effect of the decrease in $\Tvir$ at fixed $\Mhalo$. 
The left panel of Fig.~\ref{f:fV cool} also shows that $\tcoolsh/\tff$ increases with radius. This increase with radius holds even if we calculate the cooling time using the actual volume-weighted temperature in the simulation rather than using $\Tc$ (see eqn.~\ref{e:tcool}). 
An increase in $\tcoolsh/\tff$ with radius also occurs in cooling flow solutions for the CGM in which the entropy $S\equiv kT/\nH^{2/3}$ decreases inwards \citep{Stern19, Stern20a}. In contrast, isentropic CGM solutions show an opposite trend where $\tcoolsh/\tff$ decreases outwards \citep[e.g.,][]{Faerman20}\footnote{This difference can be understood by noting that at $10^5 \lesssim T \lesssim 10^7\K$ where the cooling function scales as $\Lambda\propto T^{-0.5}$ one gets $\tcool\sim kT/\nH\Lambda \propto S^{1.5}$ (neglecting metallicity gradients). In cooling flows in an isothermal potential $S\propto r$ and $\tff\propto r$, so $\tcool\propto r^{1.5}$ and hence $\tcool/\tff\propto r^{0.5}$ increases outwards. In contrast isentropic models give $\tcool\propto r^0$ and $\tcool/\tff\propto r^{-1}$, i.e.~this ratio decreases outwards.}. 

The right panel of Figure~\ref{f:fV cool} compares $\fv$ with $\tcoolsh/\tff$ in our simulations, which span a range of over $10^3$ in $\tcoolsh/\tff$. 
\changed{
The cool gas filling fraction shows a strong trend with $\tcoolsh/\tff$, decreasing from $\fv\approx0.6$ at $\tcoolsh/\tff\approx0.1$ to $\fv\lesssim0.03$ at $\tcoolsh/\tff\approx10$.
} 
We emphasize that this relation is non-trivial, since $\tcoolsh$ is independent of the actual gas temperature in the simulation (eqn.~\ref{e:tcool}). 
The panel shows that the median relations between $\fv$ and $\tcoolsh/\tff$ are almost independent of redshift and radius, suggesting that the volume filling fraction of cool gas is determined mainly by $\tcoolsh/\tff$ rather than by other system parameters. 

In Figure~\ref{f:Rsonic} we plot the radius where $\tcoolsh=\tff$ as a function of $\Mvir$ for different redshifts. In cooling flow solutions this radius corresponds to the sonic point, since the Mach number of the flow roughly equals $(\tcoolsh/\tff)^{-1}$ \citep{Mathews78, Stern20a}. 
The figure shows that $R(\tcoolsh=\tff)$ tends to be beyond $0.5\Rvir$ in halos with mass $\lesssim10^{11}\msun$, while it drops to disc scales ($\lesssim0.1\Rvir$, see below) at $\Mhalo\gtrsim10^{12}\msun$. At intermediate masses ($\sim 10^{11}-10^{12}\msun$) we find $\tcoolsh=\tff$ between $0.1\rvir$ and $0.5\Rvir$, and $\fv$ is large in the inner CGM while it is small in the outer CGM. 
\changed{
Fig.~\ref{f:Rsonic} thus demonstrates the \citetalias{Stern20b} result that as FIRE halos grow in mass, $\fv$ drops and the hot phase becomes prevalent first in the outer CGM and only later in the inner CGM. This `outside-in' formation of a stable hot CGM phase is opposite to the scenario suggested by \cite{Birnboim03} based on idealized simulations, where an expanding accretion shock causes the hot phase to form from the inside out. 
}




\subsection{Self-shielding radius}\label{s:density}

\begin{figure}
\includegraphics{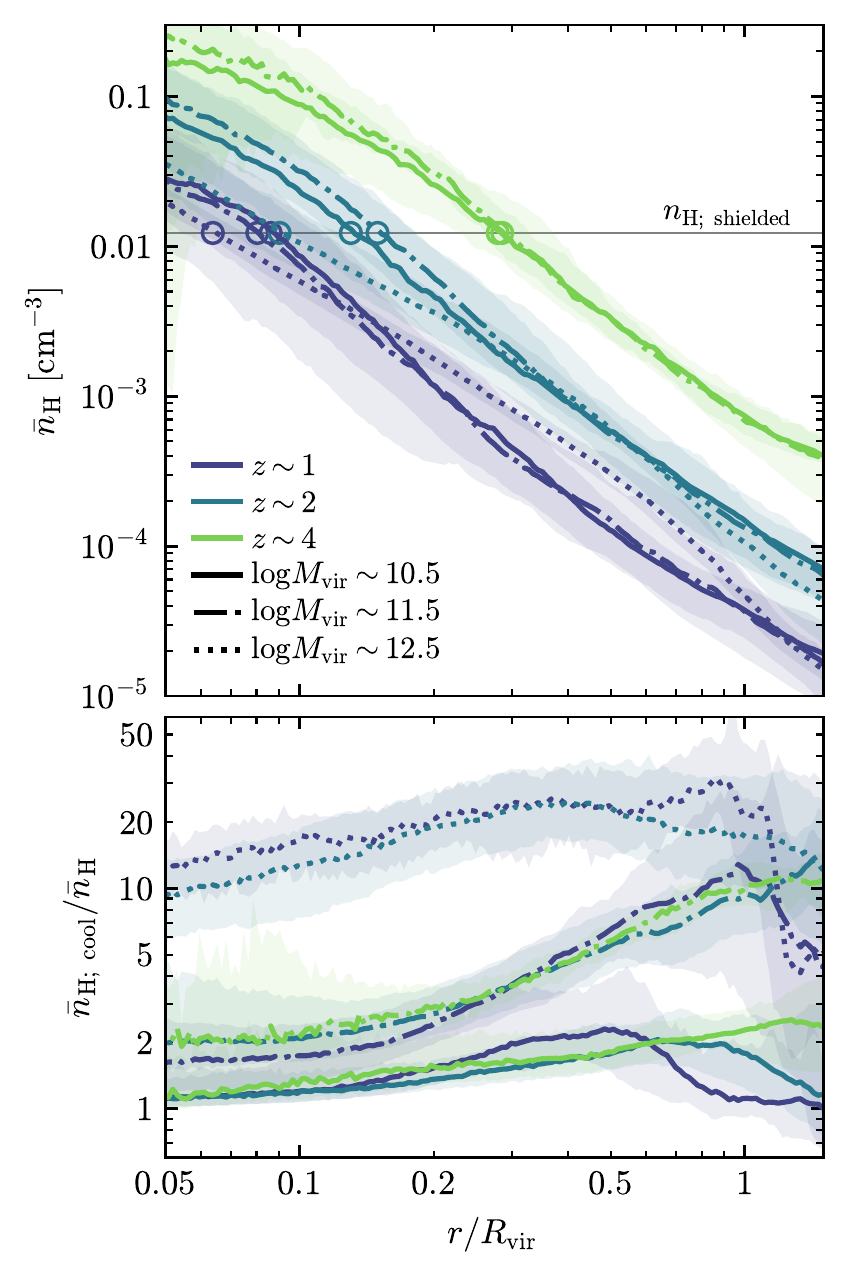}
\caption{
\textit{Top:} Spherically-averaged gas density profiles in FIRE, as a function of redshift and halo mass. For each bin in ($z$, $\Mvir$) a line and band denote the median and $\pm1\sigma$ range of density profiles among all snapshots in the bin. Note that densities increase with redshift and are roughly independent of $\Mvir$. Circles mark where the average densities exceed the shielding density, calculated using eqn.~(\ref{e:nhss}) with $\Gamma=10^{-12}\s^{-1}$ and $T=10^4\K$. 
\textit{Bottom:} The ratio of the average cool ($T<10^{4.5}\K$) gas density to the total density. The cool gas and total gas densities are comparable in $\sim10^{10.5}\msun$ halos and at small radii in $\sim10^{11.5}\msun$ halos, \changed{indicating no compression by an ambient hot phase.}
}
\label{f:density}
\end{figure}

\newcommand{\fhi}{f_{\rm HI}}

\begin{figure}
\includegraphics{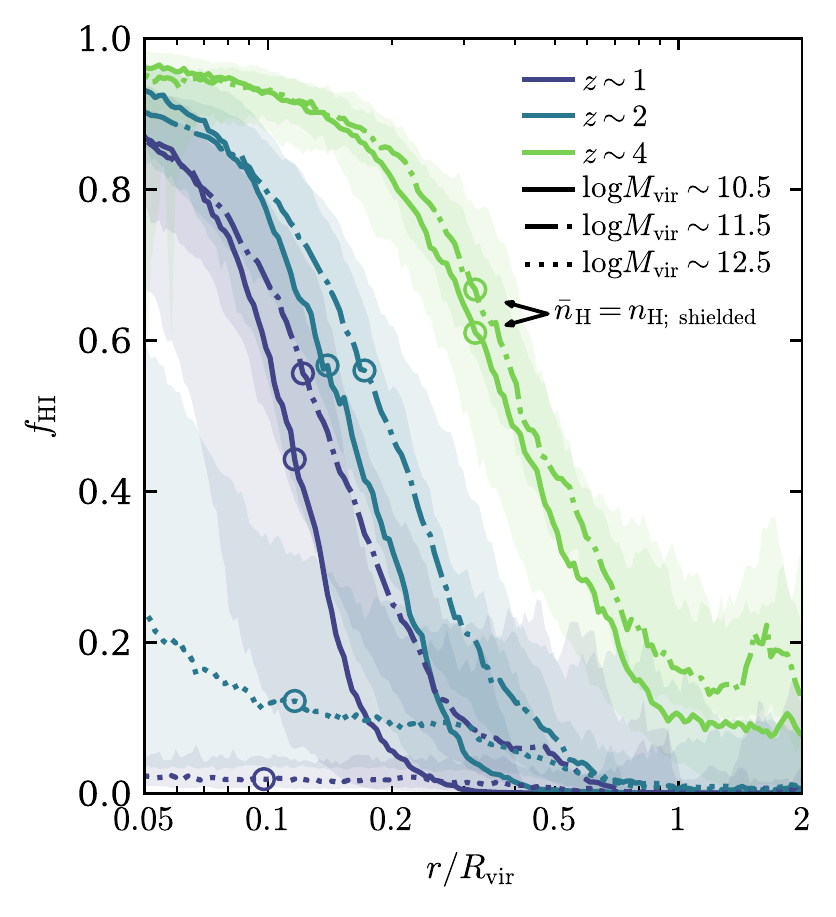}
\caption{
Neutral hydrogen fraction in the FIRE CGM, as a function of redshift and halo mass. For each ($z$, $\Mvir$) bin a line and band denote the median $\fhi$ profile and $\pm1\sigma$ range among all snapshots in the bin. 
The inner CGM is predominantly neutral in $\sim10^{10.5}\msun$ and $\sim10^{11.5}\msun$ halos, and the extent of this neutral region increases with redshift. The radius where $\fhi\approx0.5$ in $<10^{12}\msun$ halos roughly corresponds to where the mean density exceeds the shielding density (marked by circles, see Fig.~\ref{f:density}). In $>10^{12}\msun$ halos the CGM is predominantly ionized even at radii where $\nHavg>\nhss$, due to the dominance of the \changed{collisionally-ionized} hot phase. 
}
\label{f:fHI}
\end{figure}

\begin{figure}
\includegraphics{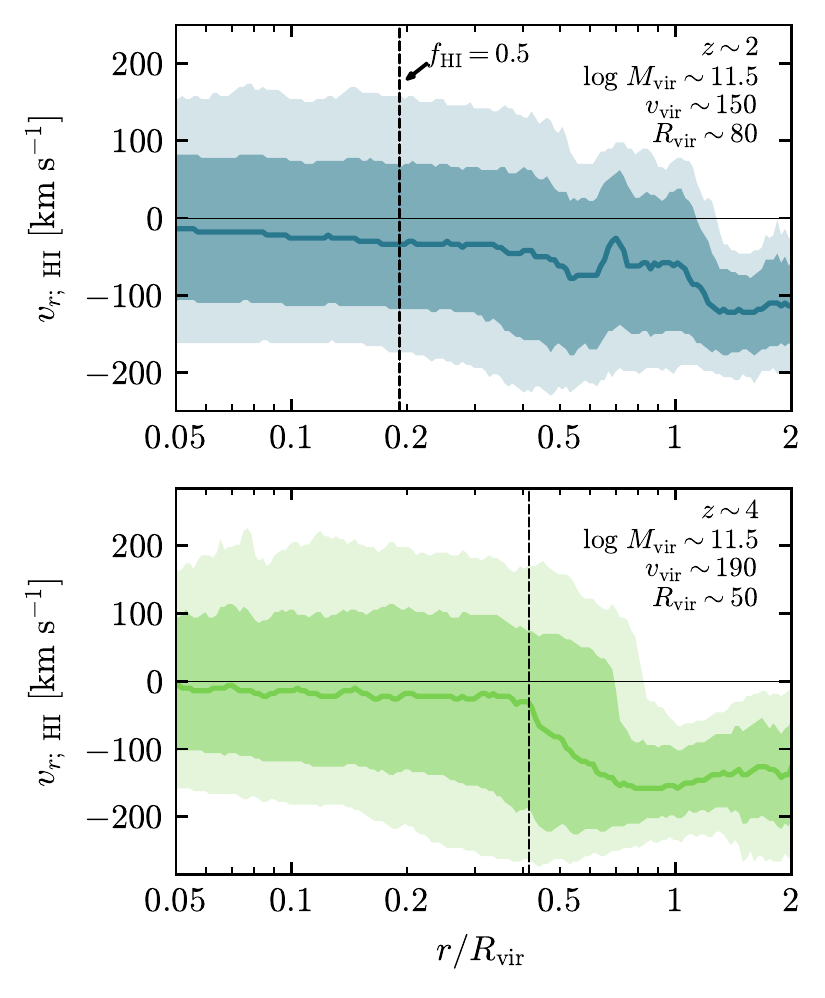}
\caption{Radial velocity distribution of the neutral CGM in FIRE. Lines mark the median $v_r$ (weighted by \hi-mass) at different radii, while colored bands denote $16-84$ and $5-95$ percentiles. The combined distribution of all snapshots in the $(z\sim2,\Mvir\approx10^{11.5}\msun$) bin is shown in the top panel, and similarly for the $(z\sim4,\Mvir\approx10^{11.5}\msun)$ bin in the bottom panel. Vertical lines mark the radius within which the CGM is predominantly neutral. Characteristic $\vvir$ and $\Rvir$ are noted in the legend. The radial velocities in the neutral inner CGM span $\pm\vvir$, \changed{with a small net inflow}.
}
\label{f:vr}
\end{figure}

\begin{figure*}
\includegraphics[width=\textwidth]{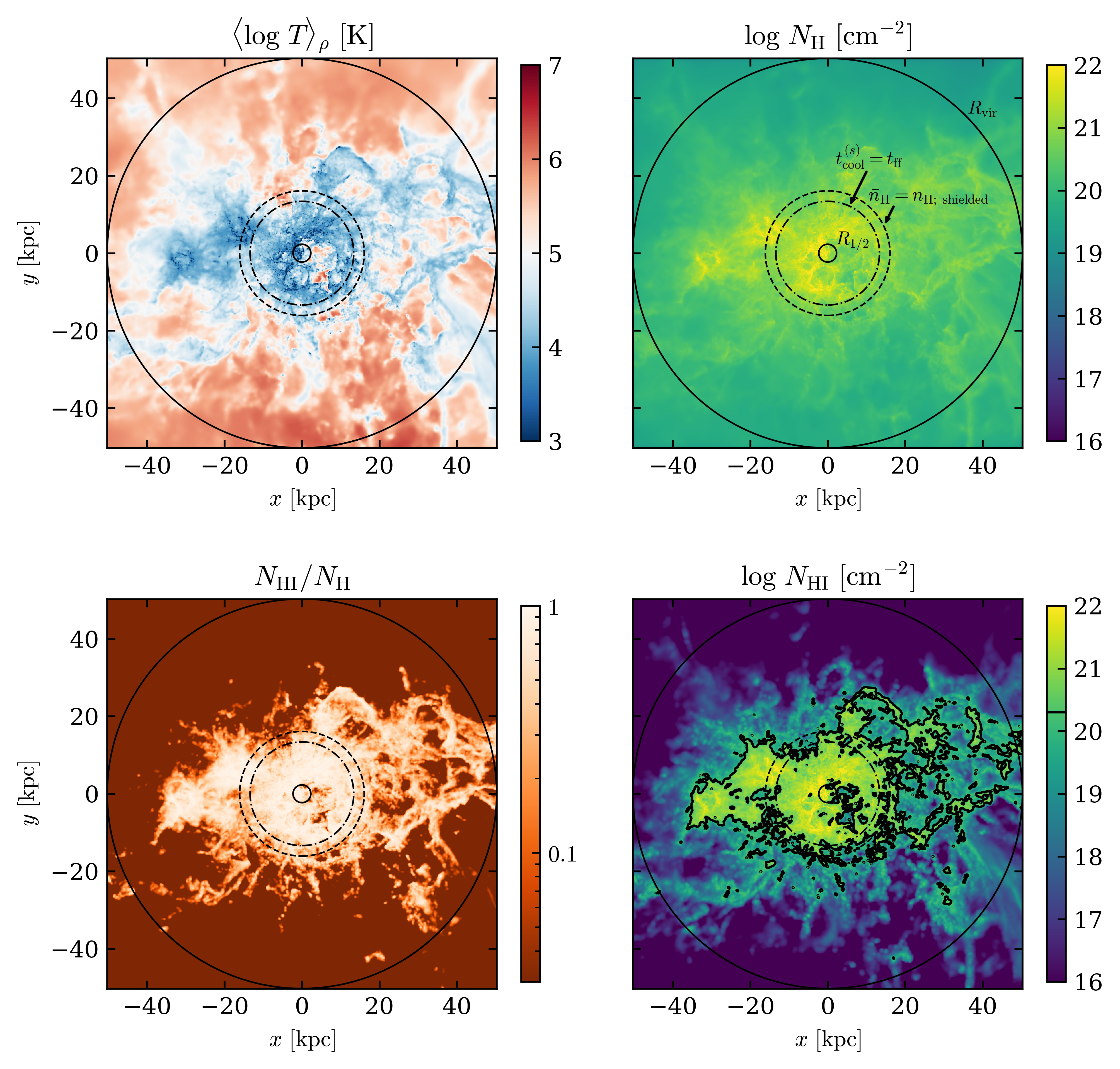}
\caption{Projected properties of a $z=4$ CGM in the m13A4 simulation. Halo and stellar mass are respectively $5\cdot10^{11}\msun$ and $0.8\cdot10^{10}\msun$. 
The panels show mass-weighted logarithmic temperature (\textbf{top-left}), hydrogen column (\textbf{top-right}), neutral hydrogen fraction (\textbf{bottom-left}), and neutral hydrogen column (\textbf{bottom-right}). Black circles mark the virial radius $\Rvir=50\kpc$, the shielding radius at which $\nHavg=\nhss$ ($16\kpc$, eqn.~\ref{e:nhss}), the radius at which $\tcoolsh=\tff$ ($13\kpc$, eqns.~\ref{e:tff}--\ref{e:tcool}), and the stellar half mass radius $\Rhalf=2.3\kpc$. 
To minimize the cross-section of the central galaxy, orientation of the images is \emph{edge-on} with respect to the angular momentum of gas at the center. DLA sightlines ($\NHI>2\cdot10^{20}\cm^{-2}$) are delineated by black contours in the bottom-right panel. 
Note that at radii smaller than $R(\tcoolsh=\tff)$ the CGM is predominantly cool and neutral, and has a DLA covering factor close to unity. 
 }
 \label{f:maps}
\end{figure*}

\begin{figure*}
\includegraphics[width=\textwidth]{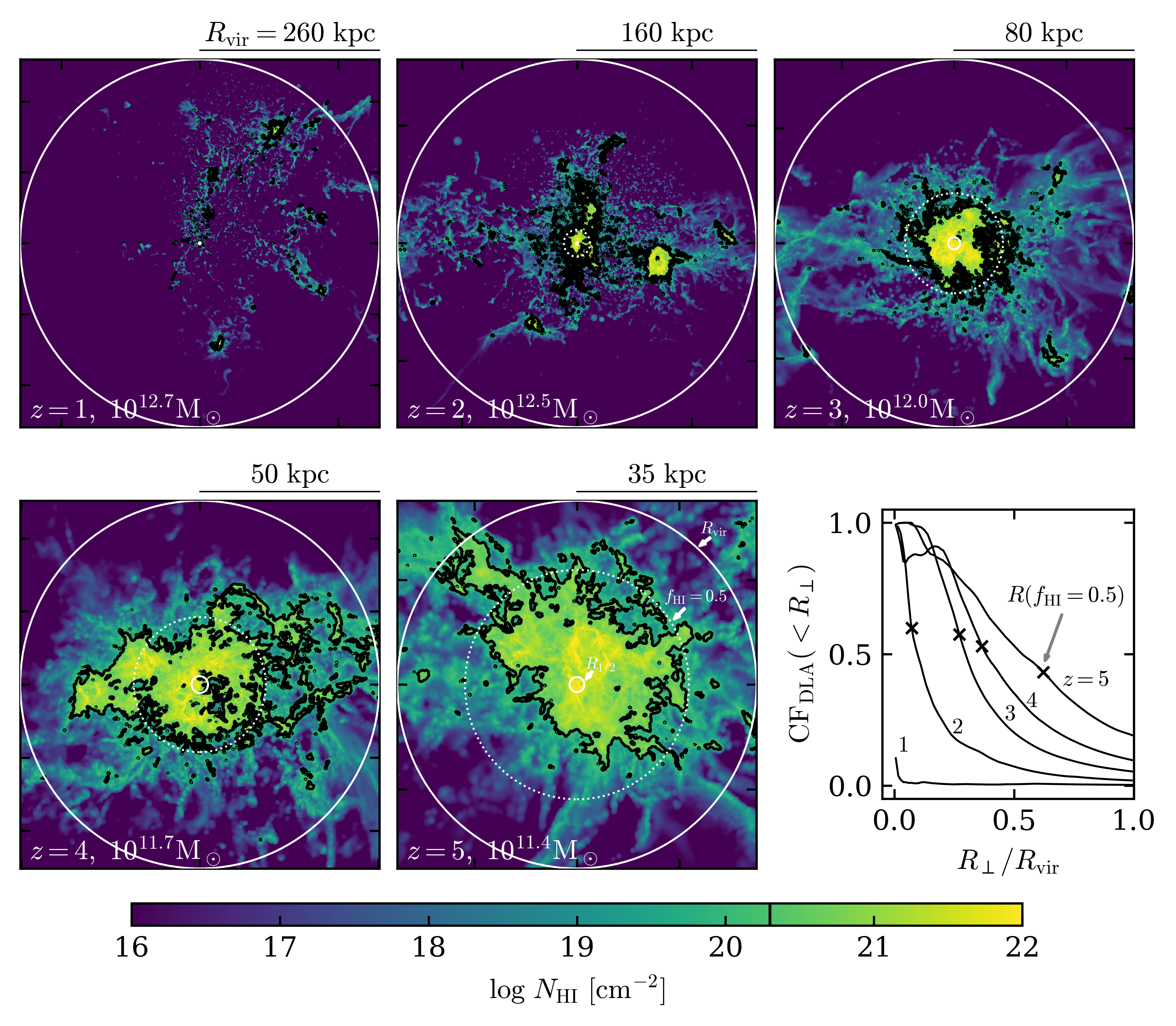}
\caption{\textbf{(Images)} \hi-columns in the m13A4 simulation as a function of redshift. The redshift and halo mass are noted in the panels. Solid white circles mark $\Rvir$ and $\Rhalf$, while dotted circles mark the radius where $\fhi=0.5$. Black contours delineate DLA sightlines.
As in Fig.~\ref{f:maps}, the projection axis is chosen so the angular momentum vector is pointed upward. 
\textbf{(Bottom-right)} The DLA covering fraction versus impact parameter for the five snapshots.
Note how the region where $\fDLA$ approaches unity expands with increasing redshift, tracing the region where the CGM is predominantly neutral. 
}
 \label{f:maps by z}
\end{figure*}

\begin{figure*}
\includegraphics{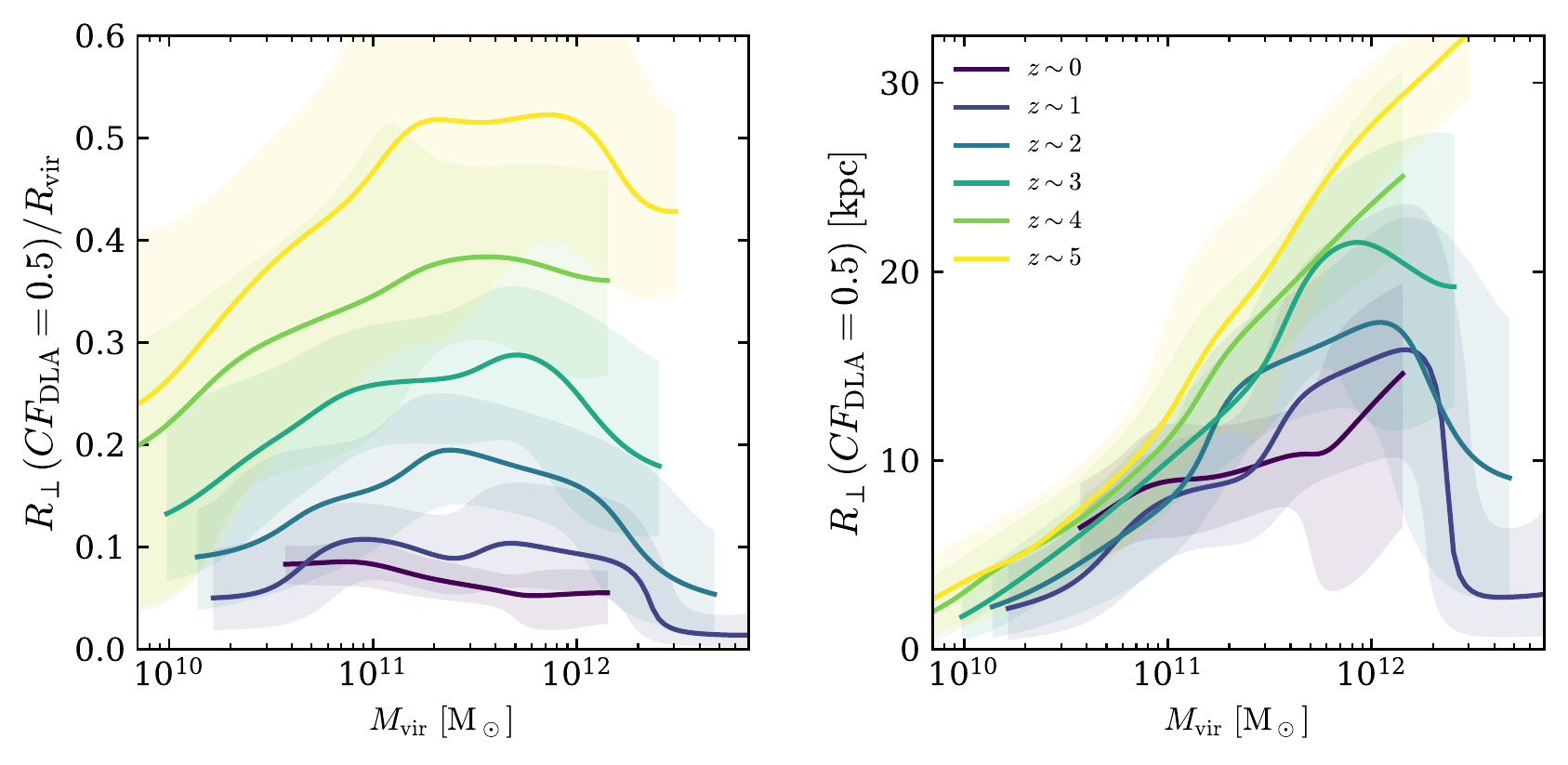}
\caption{The extent of DLAs in FIRE versus halo mass and redshift, in units of $\Rvir$ (\textit{left}) and in physical units (\textit{right}). 
Colored lines mark median relations for different redshift bins. The ratio $\Rimp(\fDLA=0.5)/\rvir$ shows a strong trend with redshift, due to mean densities exceeding $\nhss$ at larger fractions of $\rvir$ with increasing redshift (Fig.~\ref{f:density}). 
At $\Mvir\gtrsim10^{12}\msun$ the value of $\Rimp(\fDLA=0.5)$ decreases relative to lower $\Mvir$ due to the decrease in the filling factor of cool gas (Fig.~\ref{f:fV cool Mhalo}). 
} 
\label{f:ratio}
\end{figure*}

In the previous section we demonstrated that a large fraction of the inner CGM volume is cool in $\lesssim10^{12}\msun$ FIRE halos, where the cooling times of hot gas are short. Here we explore under which conditions this volume-filling cool gas is also shielded from photoionization.

As mean CGM densities at a given $r/\Rvir$ are expected to scale with cosmic mass density and be roughly independent of halo mass (see eqn.~\ref{e:nH} in the appendix), we expect the importance of shielding in a volume-filling cool phase to mainly depend on redshift. This is demonstrated in the top panel of Figure~\ref{f:density}, where we compare spherically-averaged hydrogen density profiles $\nHavg(r)$ in FIRE with the minimum density for shielding in FIRE $\nhss$. The value of $\nhss$ is estimated from eqn.~(\ref{e:nhss}) assuming $T\approx10^4\K$ and $\Gamma=10^{-12}\s^{-1}$, as measured for the UVB at $z\sim1-5$ (e.g.,~\citealt{FaucherGiguere20}). 
To derive $\nHavg(r)$ we divide each snapshot into radial shells of width $0.05\dex$. We then sum the number of hydrogen particles in all resolution elements with centers within the shell and divide by the shell volume. 
Solid lines and colored bands in Fig.~\ref{f:density} mark the median and $\pm1\sigma$ range of these spherically-averaged profiles, among all snapshots in the relevant ($z$, $\Mvir$) bin. The median densities decrease with radius as $ r^{-2.1}-r^{-2.3}$ and increase with redshift roughly as $\sim(1+z)^3$. 
The radius where $\nHavg$ exceeds $\nhss$ increases from $\approx0.05-0.1\Rvir$ at $z\sim1$ to $\approx0.3\Rvir$ at $z=4$, and is roughly independent of halo mass.

The bottom panel of Fig.~\ref{f:density} plots the ratio of the spherically-averaged density of $T<10^{4.5}\K$ gas, $\nHcool$, to the total gas density plotted in the top panel.
\changed{
The panel shows that $\nHcool\gtrsim10\nHavg$ in $10^{12.5}\msun$ halos and $\nHcool\gtrsim5\nHavg$ beyond $0.5\Rvir$ in $10^{11.5}\msun$ halos, consistent with compression of cool gas by an ambient hot phase. 
}
In contrast, $\nHcool\sim\nHavg$ in $10^{10.5}\msun$ halos and at small radii in $10^{11.5}\msun$ halos. These halos and radii correspond to regions where $\tcoolsh<\tff$ and hence the hot phase is largely absent (Fig.~\ref{f:Rsonic}), so $\nHavg$ roughly traces the density of the cool phase. 

The radius where $\nHavg=\nhss$ can also be estimated analytically as a fraction of $\Rvir$ (appendix~\ref{a:NH})
\begin{equation}\label{e:Rshielded}
 R(\nHavg=\nhss) = 0.33 \fg^{1/2}\left(\frac{1+z}{5}\right)^{3/2}\Gamma_{-12}^{-1/3}  T_4^{-0.09} \Rvir ~,
\end{equation}
and in physical units
\begin{equation}\label{e:Rshielded physical}
R(\nHavg=\nhss) = 21 \fg^{1/2}\left(\frac{1+z}{5}\right)^{1/2}\Gamma_{-12}^{-1/3}  T_4^{-0.09} M_{12}^{1/3} \kpc ~,
\end{equation}
where $\fg$ is the CGM mass in units of the halo baryon budget and $M_{12}\equiv\Mvir/10^{12}\msun$. 
Taking $\Gamma$ appropriate for the UVB in FIRE (i.e.~neglecting ionization from local stars), and calculating $\fg$ based on the total gas mass at $0.1-1\Rvir$ in the snapshots, we find that the median value of the product $\fg^{1/2}\Gamma_{-12}^{-1/3}$ somewhat decreases with redshift, from $\approx1.5$ at $z=0$ to $\approx0.9$ at $z=4$. Using these values in eqn.~(\ref{e:Rshielded}) demonstrates that it provides a good approximation for the radius at which $\nHavg=\nhss$ shown in Fig.~\ref{f:density}.

The implications of rapid cooling and self-shielding for the hydrogen neutral fraction $\fhi$ are shown in Figure~\ref{f:fHI}. To calculate $\fhi$ \changed{as a function of $r/\Rvir$} in a given snapshot we divide the gas into $0.05\dex$-wide shells in radius, and divide the \hi\ mass by the total hydrogen mass within each shell. For each bin of ($z$, $\Mvir$) we then calculate the median and $\pm1\sigma$ range of the $\fhi$ profiles among all snapshots in the bin. Fig.~\ref{f:fHI} shows that the inner CGM has a high neutral fraction in $<10^{12}\msun$ halos, while the extent of this neutral region increases with redshift. On average we find $\fhi>0.5$ out to $0.1-0.15\Rvir$ at $z=1$, out to $\approx0.2\Rvir$ at $z=2$, and out to $\approx0.4\Rvir$ at $z=4$. These radii are similar in the low-mass ($10^{10.5}\msun$) and intermediate-mass ($10^{11.5}\msun$) bins, and are comparable to the radii at which the mean gas density exceeds $\nhss$ (marked by circles). In contrast, in the high mass bin ($10^{12.5}\msun$) the CGM is predominantly ionized at all radii, due to the collisionally-ionized hot phase filling the halo when $\tcoolsh>\tff$.

\subsection{Neutral CGM kinematics}

Figure~\ref{f:vr} explores the kinematics of neutral inner CGM in FIRE. For each radial shell, we measure the distribution of radial velocities of resolution elements within the shell, combining shells with the same $r/\Rvir$ from all snapshots in a given ($\Mvir,z$) bin. To focus on the neutral CGM, we weigh each resolution element by its \hi-mass. The top panel shows the distributions in the $(z\sim2,\Mvir\approx10^{11.5}\msun$) bin while the bottom panel shows the distributions in the $(z\sim4,\Mvir\approx10^{11.5}\msun$) bin. Vertical lines mark the median radius at which $\fhi=0.5$ (Fig.~\ref{f:fHI}), which increases with redshift from $\approx0.2\Rvir\approx16\kpc$ at $z\sim2$ to $\approx0.4\Rvir\approx20\kpc$ at $z\sim4$. Lines mark the median of the distribution while colored bands denote the $16-84$ and $5-95$ percentiles. 

\changed{
Fig.~\ref{f:vr} demonstrates that at radii where the CGM is predominantly neutral (left of the dashed lines) the $5-95$ percentile range of $v_{r, \hi}$ spans $\pm150\kms$ in the $z\sim2$ panel and $\pm200\kms$ in the $z\sim4$ panel, roughly equal to the median virial velocity $\vvir$ of snapshots in the bin (noted in the legends).
The \hi-weighted median velocity is negative (inflow), though with relatively small absolute values of $20 - 30\kms$. The mean of the distribution (not shown) is similar to the median. Other redshifts and masses in our sample exhibit a similar spread of $-\vvir \lesssim v_{r; \hi}\lesssim\vvir$ and a small negative median $v_{r; \hi}$. 

While Fig.~\ref{f:vr} demonstrates that the average inflow velocity in inner neutral CGM is small, we emphasize that the CGM is not thermal-pressure supported since there is no ambient hot phase with temperature $\sim\Tvir$. Neutral CGM in FIRE more resemble `dynamically-supported CGM', where support against gravity is provided by bulk flows and turbulence 
arising from the interaction between cosmological inflows and outflowing galactic winds, as seen in idealized CGM simulations with $\tcoolsh<\tff$ \citep{Fielding17}. 
}

\section{The DLA covering factor of the CGM in FIRE}\label{s:CF}

Above we demonstrated that FIRE halos have a high neutral fraction at CGM radii where $\tcoolsh<\tff$ and $\nH>\nhss$. We now discuss the implications for DLA covering factors.

Figure~\ref{f:maps} plots projected quantities in the $z=4$ snapshot of the m13A4 simulation. The halo and stellar masses in this snapshot are $\Mhalo=5\cdot10^{11}\msun$ and $\Mstar=0.8\cdot10^{10}\msun$, and the galaxy star formation rate is ${\rm SFR}=16\msun\yr^{-1}$. 
The top-left panel shows the mass-weighted logarithmic temperature $\langle\log T\rangle_\rho$, derived by depositing the temperature of particles on an evenly-spaced grid according to their smoothing kernel, and then averaging over $\pm\Rvir$ in the direction perpendicular to the image. 
The other panels show similar projection maps of $\NH$, $\NHI/\NH$ and $\NHI$. To focus on the CGM and minimize the effects of an extended gaseous disc we choose an edge-on orientation, such that the total angular momentum vector of gas in the inner $0.05\Rvir$ is directed upward. We also mark in the panels the virial radius $\Rvir=50\kpc$, the stellar half mass radius $R_{1/2}=2.3\kpc$ (measured using all stars within $0.15\Rvir$), the radius where $\nHavg=\nhss$ ($16\kpc$), and the radius where $\tcoolsh=\tff$ ($13\kpc$). 

The top-left panel of Fig.~\ref{f:maps} shows that at radii smaller than $R(\tcoolsh=\tff)$, the CGM is predominantly cool with temperatures $10^3-10^4\K$, orders of magnitude lower than the halo virial temperature of $\Tvir=1.5\cdot10^6\K$. The absence of a dominant hot phase is consistent with rapid cooling within this radius. The bottom-left panel shows that this inner region is also neutral along practically all sightlines, as expected since typical densities are above $\nhss$. In contrast, at larger radii cool temperatures and high neutral fractions are seen only along specific angles. 

\changed{
The top-right panel in Fig.~\ref{f:maps} shows that $\NH$ exceeds $10^{20}\cm^{-2}$ in the inner CGM of this halo, typically by a large factor of $\gtrsim10$, as expected in halos with mass $\Mhalo\sim10^9\msun$ or larger (see appendix~\ref{a:NH}). 
}
A large neutral fraction in the inner CGM is hence a sufficient requirement for a large DLA covering factor in such halos, as evident in the bottom-right panel. We find a DLA covering factor of $\fDLA=0.75$ at impact parameters $\Rimp<R(\tcoolsh=\tff)$ in the edge-on projection shown in Fig.~\ref{f:maps}, and a similar $\fDLA=0.74$ at the same impact parameters in a face-on projection. 

Figure~\ref{f:maps by z} plots $\NHI$ maps (similar to the bottom-right panel of Fig.~\ref{f:maps}) for snapshots of the m13A4 simulation in the range $1\leq z \leq 5$. As in Fig.~\ref{f:maps} the projection axis is chosen so the angular momentum vector of gas at the center is pointed upward in all images. 
A strong increase in the DLA covering fraction of the CGM with redshift is apparent, from a few small patches of DLA sightlines at $z=1$ to a DLA covering fraction approaching unity out to $\sim0.5\rvir$ at $z=5$. This trend is quantified in the bottom-right panel, where we plot $\fDLA(<\Rimp)$ versus $\Rimp/\rvir$ for each of the five snapshots. This panel also shows that the impact parameter where $\fDLA=0.5$ roughly corresponds to the 3D radius where $\fhi=0.5$ (marked by crosses). 


Figure~\ref{f:ratio} plots the dependence of the DLA covering factor on halo mass and redshift in all 16 FIRE simulations. The left panel plots $\Rimp(\fDLA=0.5)/\Rvir$, where $\Rimp(\fDLA=0.5)$ is defined as the impact parameter within which $50\%$ of sightlines have a DLA. We calculate $\Rimp(\fDLA=0.5)$ in a random edge-on projection of each simulation snapshot, and use colored lines and bands to denote the median and dispersion of all snapshots in different redshift bins. The right panel plots $\Rimp(\fDLA=0.5)$ in physical units. 

The left panel of Fig.~\ref{f:ratio} demonstrates that the DLA covering fraction of the CGM increases strongly with redshift. Specifically, in $10^{11}-10^{12}\msun$ halos the value of $\Rimp(\fDLA=0.5)/\Rvir$ increases by an order of magnitude from $\approx0.05$ at $z=0$ to $\approx0.5$ at $z=5$. The ratio $\Rimp(\fDLA=0.5)/\rvir$ also exhibits a weak dependence on halo mass at $10^{10.5}-10^{12}\msun$. Both these trends are expected from the relation between average CGM densities and $\nhss$ shown in Fig.~\ref{f:density} and eqn.~(\ref{e:Rshielded}), which show that $R(\nHavg=\nhss)/\rvir$ increases strongly with redshift and is roughly independent of halo mass. The right panel shows that in physical units $\Rimp(\fDLA=0.5)$ depends only weakly on redshift, increasing by at most a factor of two from $z=0$ to $z=5$ for a given halo mass, and increases with halo mass up to $\Mhalo\approx10^{12}\msun$, as expected from eqn.~(\ref{e:Rshielded physical}) . At larger halo masses $\Rimp(\fDLA=0.5)$ drops, as most evident in the $z\sim1$ and $z\sim2$ bins. This drop is due to the high $\tcoolsh/\tff$ in the inner CGM of these halos, which implies a small $\fcool$ and hence a low neutral fraction (Fig.~\ref{f:fHI}). 

We note that the weak dependence of $\Rimp(\fDLA=0.5)$ on redshift in physical units (right panel of Fig.~\ref{f:ratio}), is a result of the relatively weak dependence of mean CGM densities on redshift at fixed physical radius and halo mass, so the physical radius where $\nHavg=\nhss\approx0.01\cm^{-3}$ weakly depends on redshift. This is in contrast with the strong evolution of $\Rimp(\fDLA=0.5)$ with redshift in units of $\Rvir$ (left panel of Fig.~\ref{f:ratio}). This behavior is phenomenologically similar to the lack of evolution of the inner DM profile in physical units \citep[e.g.,][]{Diemer13,WetzelNagai15}, a similarity which warrants future investigation. 

\section{Comparison to observations}\label{s:obs}


%


\changed{
In this section we compare neutral inner CGM in FIRE with observations of DLA -- galaxy pairs at $z\sim4$. To this end, we calculate the predictions of FIRE for the impact parameters, metallicities, velocity dispersions, and number densities of $z\sim4$ DLAs. 
}

\subsection{DLA metallicities}
 
\begin{figure}
\includegraphics{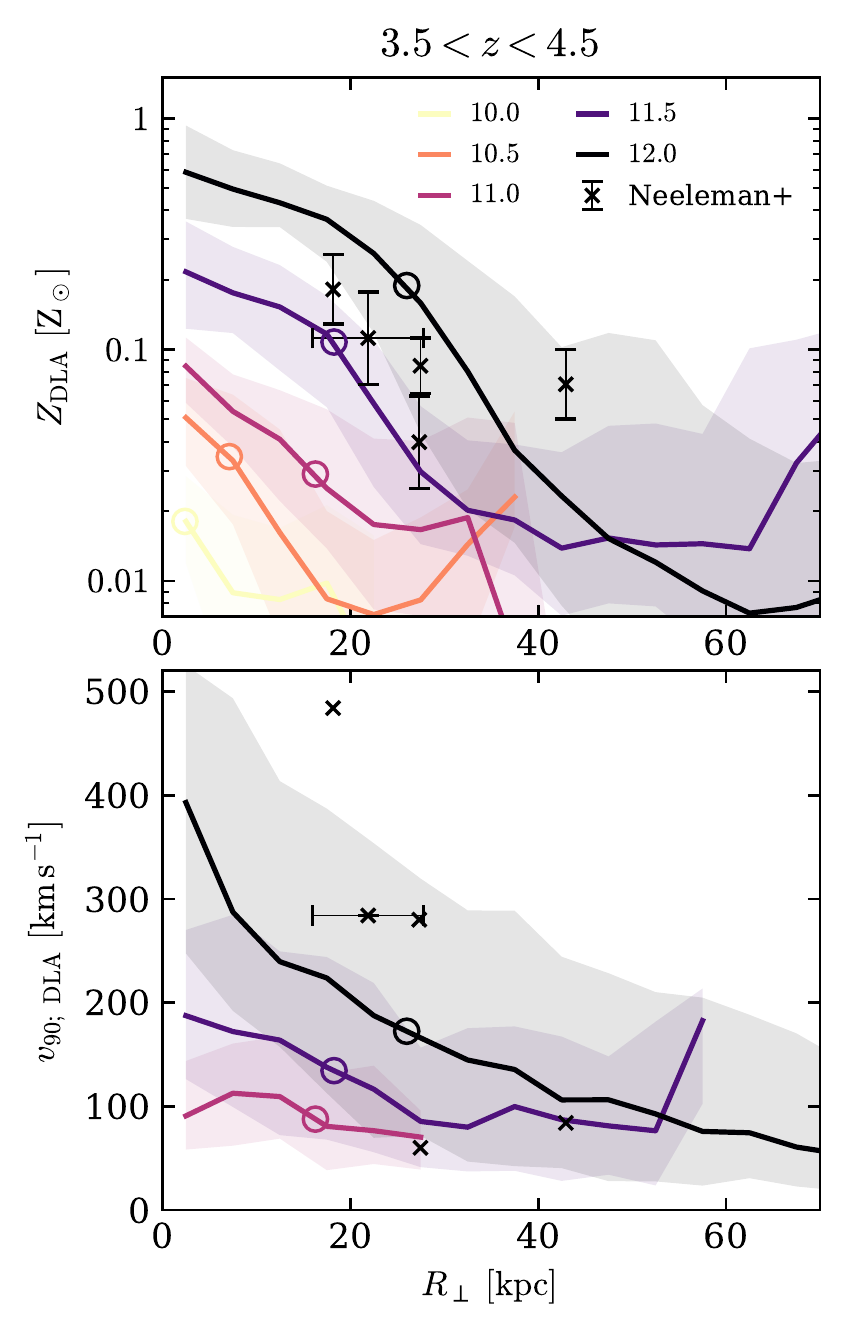}
\caption{
DLA metallicities (\textbf{top}) and velocity widths (\textbf{bottom}) in $z\sim4$ FIRE halos. 
Solid lines and colored bands denote medians and $16-84$ percentile ranges, as a function of impact parameter and halo mass (noted in the legend in $\log\msun$). 
Circles mark the radii within which the DLA covering factor is $50\%$ (typically $\approx0.4\Rvir$, see Fig.~\ref{f:ratio}). 
Error-bars mark impact parameters between high-metallicity DLAs and bright galaxies measured by \citet{Neeleman17,Neeleman19}. 
\changed{
The observed DLA impact parameters, metallicities, and velocities fall within the $16-84$ percentile ranges of $10^{11.5}-10^{12}\msun$ halos in FIRE.
}
}
\label{f:Z}
 \end{figure}

In the top panel of Figure~\ref{f:Z} we plot DLA metallicity $\ZDLA$ in FIRE as a function of impact parameter and halo mass. 
To derive $\ZDLA(\Mhalo,\Rimp)$ we use all $3.5<z<4.5$ snapshots from the 16 FIRE simulations. 
We choose $50$ randomly-oriented mock sightlines with $\NHI>2\cdot10^{20}\cm^{-2}$ through each snapshot, and calculate the average metallicity of each sightline, weighted by the \hi-mass of resolution elements along the sightline. As above we include particles within $\pm\Rvir$ from the mid-plane in the averaging. 
The sightlines are then binned by $\Rimp$ and by $\Mhalo$ of the snapshot, and in Fig.~\ref{f:Z} we plot the median and $16-84$ percentile range of each ($\Mhalo$, $\Rimp$) bin. 
The figure shows an increase of $\ZDLA$ with halo mass, and 
a relatively flat metallicity profile out to $\approx20\kpc\approx0.3-0.5\rvir$ in the two most massive bins. This suggests a relation between mass and metallicity in the inner CGM similar to that in the ISM. 

We compare the predicted CGM metallicity profiles to the \cite{Neeleman17,Neeleman19} observations of DLA--galaxy pairs. The \citeauthor{Neeleman17} sample is based on a selection of six high-metallicity DLAs ($\ZDLA=0.05-0.2\zsun$) from a larger sample of DLAs along random quasar sightlines \citep{Rafelski12}, that were followed up with ALMA observations. 
\cii\ $158\mu{\rm m}$ emission was detected around five of the six DLAs at impact parameters $20-40\kpc$, 
potentially originating from the associated galaxy. Around one of the DLAs two \cii\ sources were detected, so the associated galaxy in this object is uncertain. Four of the six \cii\ sources also have detected continuum infrared emission, while optical emission was not detected in any object (potentially due to dust obscuration). The galaxy SFRs inferred by \citeauthor{Neeleman17} from the infrared luminosities are $7, 15, 24,$ and $110\msun\yr^{-1}$. These SFRs are similar to those of luminous Lyman-break galaxies, and to the SFRs in the central galaxy of the four massive ($\Mvir\sim10^{11.5}-10^{12}\msun$) halos in our sample, which have $300\Myr$-averaged SFRs of $2.5, 15, 75,$ and $78\msun\yr^{-1}$ at $z=4$ (see figure~15 in \citetalias{Stern20b}).

Error bars in the top panel of Fig.~\ref{f:Z} plot metallicity and impact parameter measurements from the observed sample. In the object with two possible associated galaxies we plot a horizontal error bar to denote the uncertainty in $\Rimp$. 
\changed{
The figure shows that the metallicities and impact parameters in the \citeauthor{Neeleman19} sample fall within the $\pm1\sigma$ range of $10^{11.5}-10^{12}\msun$ halos in FIRE, the same halo mass range suggested by the observed SFR.
}

Four of the five \citeauthor{Neeleman19}\ DLA--galaxy pairs have $\Rimp\lesssim30\kpc$, while $1-2$ objects have $\Rimp\lesssim20\kpc$ (depending on the actual $\Rimp$ in the object with two possible hosts). These impact parameters are comparable to $\Rimp(\fDLA=0.5)$ in the massive halos, which equals $30\kpc$ in the $10^{12}\msun$ bin and $20\kpc$ in the $10^{11.5}\msun$ bin.
Thus, a substantial fraction of the \citeauthor{Neeleman19} sample appears to originate in the region where the DLA covering factor \changed{is predicted to be close to unity}, i.e.~where the CGM is predominantly cool and neutral. 

\subsection{DLA kinematics}

The bottom panel of Fig.~\ref{f:Z} plots predicted DLA velocity widths, $\vDLA$. To estimate $\vDLA$ in the simulation we create mock observations of Si~{\sc ii}~$\lambda1808$ absorption spectra of sightlines through FIRE halos, using the {\sc Trident} package \citep{Hummels17}. We  measure the velocity range which includes 90\%\ of the absorption flux in the mock spectra, excluding the outermost 5\%\ of the flux on both the red end and the blue end of the absorption spectrum. 
The range and dispersion in predicted $\vDLA$ are then calculated using the same binning in $\Mhalo$ and $\Rimp$ as in the top panel. 
The predicted $\vDLA$ in the $\Mhalo\sim10^{11.5}\msun$ mass bin (blue) can be compared to the distribution of radial velocities \changed{($v_{r, {\rm HI}}$)} shown in the bottom panel of Fig.~\ref{f:vr}. At $\Rimp=20\kpc$ ($\approx 0.4\Rvir$) we find a $16-84$ percentile range of $50\lesssim \vDLA \lesssim250\kms$, compared to $-150\lesssim v_{r, {\rm HI}}\lesssim100\kms$, indicating that often both inflowing gas and outflowing gas contribute to the absorption along a single sightline. 

\changed{
Four of the five observed $\vDLA$ are within the $16-84$ percentile range of $10^{11.5}-10^{12}\msun$ halos (the same halo mass range suggested by the DLA metallicities and galaxy SFRs). The largest observed $\vDLA=500\kms$ is in the $5-95$ percentile range of the $10^{12}\msun$ halo bin. 
}

  \begin{figure}
\includegraphics{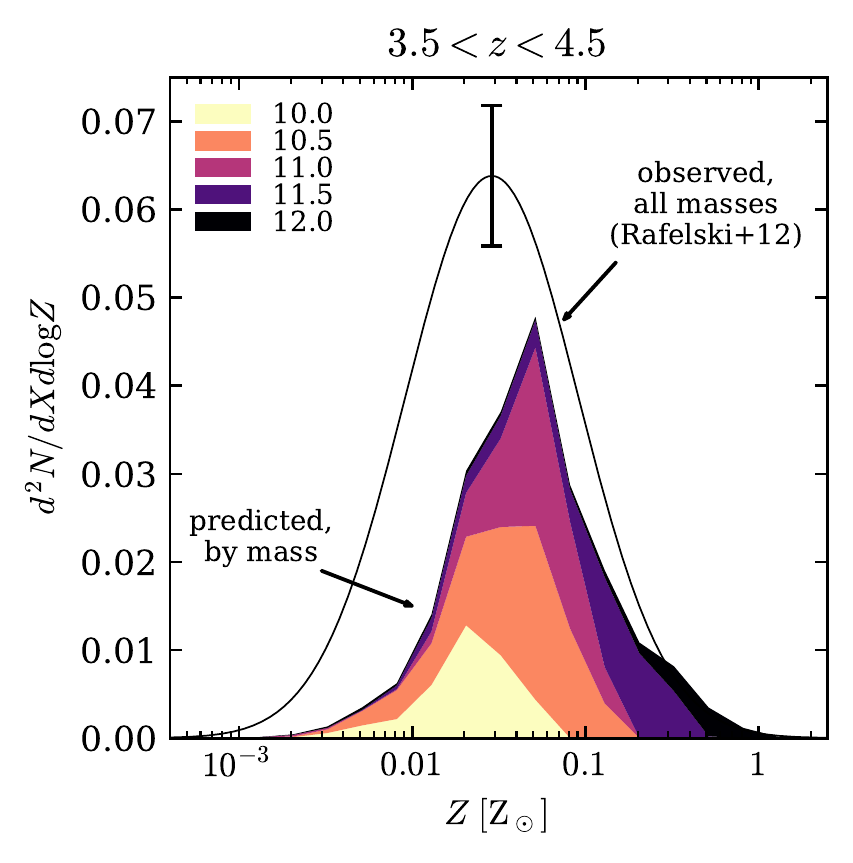}
\caption{The contribution of halos with different masses to the $z\sim4$ DLA population, as a function of DLA metallicity. Color bands plot predicted DLA number densities for different $\Mhalo$, derived from the covering area of DLAs in FIRE halos convolved with the halo number densities from the Bolshoi-Planck dark matter simulation. Note that massive halos ($\gtrsim10^{11.5}\msun$) dominate the high metallicity ($>0.1\zsun$) DLA population. The Gaussian line marks the \citet{Rafelski12} fit to the observed DLA metallicity distribution, while the error-bar denotes the uncertainty in the Gaussian normalization.
The FIRE simulations of $>10^{10}\msun$ halos reproduce the observed number of DLAs with $Z>0.05\zsun$, while the number of lower metallicity DLAs is underpredicted. 
}
\label{f:dNdXdlogZ}
\end{figure}

\subsection{DLA number densities}\label{s:number density}

The predicted number of DLAs along IGM sightlines is plotted in Figure~\ref{f:dNdXdlogZ}, separated into the contribution from different halo masses and different DLA metallicities. To estimate the expected number of DLAs we use
\begin{equation}\label{e:dN/dXdM}
 \frac{d^3\NDLA}{dX d\log\Mhalo d\log Z} = \frac{d^2 N_{\rm h}}{dXd \log \Mhalo}\fDLA(<\Rvir,\Mhalo,Z)
\end{equation}
where $\NDLA$ is the number of DLAs along the line of sight, $dX$ is the normalized pathlength \citep[e.g.,][]{Wolfe05}:
\begin{equation}
 dX  = (1+z)^3 \left|\frac{cdt}{dz}\right|\frac{H_0}{c}dz ~,
\end{equation}
and $\fDLA(<\Rvir,\Mhalo,Z)d\log Z$ is the covering factor of DLAs with metallicity in the range $(Z,Z+d\log Z)$ in FIRE halos with mass in the range $(\Mhalo, \Mhalo+d\log\Mhalo)$. We calculate the number of halos per normalized pathlength and per decade in mass using 
\begin{equation}
 \frac{d^2 N_{\rm h}}{dXd \log \Mhalo} = \frac{c}{H_0}\frac{d n_{\rm h}}{d\log\Mhalo}\pi \Rvir^2 ~,
\end{equation}
where $d n_{\rm h}/d\log\Mhalo$ is the comoving halo number density taken from \cite{Behroozi19}, which is based on the Bolshoi-Planck dark matter simulation \citep{Klypin16}. 
Fig.~\ref{f:dNdXdlogZ} demonstrates that $10^{10}-10^{11}\msun$ halos contribute a larger fraction of DLAs than more massive halos, but  high-metallicity DLAs with $Z>0.1\zsun$ are dominated by massive halos with $\Mhalo>10^{11.5}\msun$. 


The Gaussian line in Fig.~\ref{f:dNdXdlogZ} marks the observed $d^2N/dX d\log Z$ based on the analysis of \cite{Rafelski12}, which showed that observed metallicities of $z\approx4$ DLAs follow a log-normal distribution with a mean of $10^{-1.54}\zsun$ and width of $0.5\dex$. We set the normalization of the observed metallicity distribution to reproduce the integrated $dN/dX=0.08\pm0.01$ measured by \cite{ProchaskaWolfe09}, while the errorbar denotes the uncertainty in the $dN/dX$ measurement. The predictions of the FIRE simulations is consistent with the observed number of DLAs with $Z>0.05\zsun$, while the number of lower-$Z$ DLAs is underpredicted, by a factor of $\approx5$ at $Z\lesssim0.01\zsun$. 
We discuss possible origins for this difference in section~\ref{s:CGM properties} below.

%


\section{Discussion}\label{s:discussion}

\changed{
In this paper we showed that the inner CGM ($\lesssim0.4\Rvir$) of $z\sim4$ halos with mass $\lesssim10^{12}\msun$ is predominantly neutral in FIRE, leading to large DLA covering factors. We further demonstrated that the inner CGM is also neutral in lower redshift halos, though out to a smaller fraction of $\Rvir$ ($\lesssim0.2\Rvir$ at $z\sim2$ and $\lesssim0.1\Rvir$ at $z\sim1$, see Fig.~\ref{f:fHI}).
}
We showed that a high neutral fraction occurs in the CGM when two conditions are satisfied:  cooling times are sufficiently short so that hot gas cools rapidly; and densities are sufficiently high so cool gas is shielded from photoionization. 
Such a neutral CGM differs from the standard CGM picture based mainly on lower redshift observations and simulations, in which the CGM is a mixture of hot collisionally-ionized gas and cool photoionized clumps or filaments \citep[e.g.][]{Tumlinson17}. In this section we compare our results to previous studies, and discuss several uncertainties and implications. 

%


\subsection{Previous models of DLAs from the volume-filling phase of the CGM}\label{a:ZME02}


\citet[hereafter ZME02]{ZhengMiraldaEscude02} have also argued that high-redshift DLAs originate in a volume-filling neutral CGM. They assumed a spherical CGM in thermal and ionization equilibrium with the UVB, and calculated the optical depth along rays in different directions. They derived a radius $\RthickZME$ within which gas is shielded from photoionization, which scaled to our assumed cosmology and UVB equals (see appendix \ref{a:NH})
\begin{equation}\label{e:Rshielded ZME02}
 \RthickZME = 0.46 \, M_{12}^{1/9}\left(\frac{1+z}{5}\right)^{1.6} \fg^{2/3}\Gamma_{-12}^{-1/3}  \Rvir  ~.
\end{equation}
The value of $\RthickZME$ can be compared to the radius at which the average gas density equals $\nhss$ (Fig.~\ref{f:density})
and to the radius where the DLA covering factor is $50\%$ in FIRE (Fig.~\ref{f:ratio}). These three estimates are comparable both in magnitude and in dependence on halo parameters. 
Our conclusion that inner CGM are neutral at high redshift is thus quantitativley similar to the conclusion of ZME02, despite the significant difference in calculation method.  

\changed{
More recently, \cite{Theuns21} showed that a model where the CGM is filled with cool cosmological accretion is consistent with the observed column-density distribution of DLAs, and with the independence of the DLA distribution function on redshift at $z\sim2-5$. Their model is similar to our results based on FIRE in that the inner CGM is predominantly cool, 
though in FIRE the inner CGM is not pure inflow but rather a mixture of inflows and outflows (as indicated, e.g., by gas kinematics in Fig.~\ref{f:vr} and by gas metallicity in Fig.~\ref{f:Z}). The CGM profile of $\nH\propto r^{-2.2}$ assumed by \cite{Theuns21} is also similar to the slopes of $r^{-2.3}-r^{-2.1}$ in FIRE, while their CGM density normalization is roughly a factor of two lower (compare their figure~1 with Fig.~\ref{f:density} above)\footnote{\changed{The density profile assumed in \cite{Theuns21} is based on self-similar solutions for cool cosmological accretion \citep{Bertschinger85}. It is intriguiong that this calculation roughly reproduces the profiles in FIRE despite the addition of outflows from the galaxy, which warrants further investigation.}}. 
Thus, the DLA column density distribution predicted by \cite{Theuns21} for the inner CGM are expected to be similar to those in FIRE. 
}

Our results extend the analytic studies of ZME02 and \cite{Theuns21} in two main ways. 
First, we demonstrate that their assumption of a cool and quasi-spherical CGM is valid in FIRE when $\tcoolsh\lesssim\tff$ (Fig.~\ref{f:fV cool}), and present simulation results where the range in halo mass, redshift and radius in which this condition is satisfied. 
\changed{
Specifically, we find that a large volume fraction of the inner CGM is cool at halo masses $\lesssim5\cdot10^{11}\msun$ at all redshifts, while the outer CGM is cool at lower halo masses of $\lesssim 10^{11}\msun$ and mainly at $z>2$ (Fig.~\ref{f:fV cool Mhalo}). 
}
Second, we demonstrate that a neutral CGM occurs even when accounting for the complexities in the FIRE cosmological simulations neglected by these analytic studies, including deviations from spherical symmetry, heating by accretion and stellar feedback, and the contribution of winds from the central and satellite galaxies to the CGM \citep[for an analysis of these processes in FIRE see][]{Muratov15,anglesalcazar17a,Hafen19,Hafen20}.

\subsection{The extent of neutral CGM versus disk sizes}

HST observations of galaxy sizes at $z\sim4$ find effective radii $r_e\approx 0.5-2\kpc$ for stellar masses $10^9-10^{11}\msun$, corresponding to $r_e \sim 0.01-0.035\Rvir$ according to abundance matching estimates \citep{Shibuya15}. These measurements are roughly consistent\footnote{At lower redshifts $z\lesssim2$, the four galaxies in our sample with $\Mvir>10^{12}\msun$ are more compact than in observations, see \cite{Wellons20} and \cite{Parsotan21}.} with the four simulated galaxies in our sample in this stellar mass range, which have stellar half mass radii $R_{1/2}=0.8-2.3\kpc$. Our result that neutral CGM and high DLA covering factors extend to $r\approx20\kpc\approx0.4\rvir$ at this redshift (Figs.~\ref{f:fHI} and \ref{f:ratio}) thus implies that DLAs extend to radii more than an order of magnitude larger than $r_e$. 
This conclusion is consistent with previous studies which demonstrated that DLA number densities are higher than expected based on the cross-section of galactic discs \citep{Berry14, DiGioia20}. 

We find that the extent of the region where DLA covering factors are large decreases with cosmic time with respect to $\Rvir$, due to lower gas densities and hence less shielding from photoionization. Since disc sizes are also expected to scale with $\Rvir$ \citep[e.g.,][]{Mo98}, this result suggests there is a critical redshift at which DLAs transition from originating in quasi-spherical neutral CGM to originating in central discs. In our sample of FIRE galaxies angular momentum support becomes important at $\lesssim0.1\rvir$ (see figure~13 in \citetalias{Stern20b}), corresponding to the value of $\Rimp(\fDLA=0.5)$ at $z<1$ (Fig.~\ref{f:ratio}). Indeed, DLA covering factors at $z<1$ differ between face-on and edge-on orientations, typically by $\approx30\%$, in contrast with at $z\gtrsim1$ when there is no systematic difference between different orientations. 
Our results thus suggest a transition occurs in the DLA population at $z\sim1$, from being dominated by neutral CGM at higher redshifts to being dominated by central discs at lower redshifts.

\subsection{CGM metallicities in FIRE versus observations}\label{s:CGM properties}

\changed{
We find that the enrichment of the $z\sim4$ CGM in FIRE is consistent with observations of high-metallicity DLAs at the same redshift. 
This is apparent in the impact parameters between DLAs and their associated galaxies (Fig.~\ref{f:Z}), and in the number densities of high-metallicity DLAs (Fig.~\ref{f:dNdXdlogZ}).  
}
Our results, however, suggest a lack of low-metallicity DLAs (Fig.~\ref{f:dNdXdlogZ}). Some of the missing low-$Z$ DLAs could originate in $\sim10^9\msun$ halos, which are not included in the predictions shown in Fig.~\ref{f:dNdXdlogZ}. We can roughly estimate the total (metallicity-integrated) contribution of these halos by extrapolating the $\Rimp(\fDLA=0.5)$ versus halo mass relation at $z=4$ to $10^9\msun$ (see Fig.~\ref{f:ratio}), and multiplying by $d^2 N_{\rm h}/dXd\log\Mhalo$ calculated as in section~\ref{s:number density}. 
This gives $d^2N/(dXd\log\Mhalo)\approx0.02$, roughly half of the deficit in the FIRE prediction. Halos with an even lower mass are unlikely to contribute significantly to the DLA population due to CGM depletion by UVB photoheating \citep{Okamoto08,Theuns21}. Thus, the covering factors of DLAs in low-mass halos and/or at large radii where metallicities are low may be underpredicted by FIRE. Larger samples of DLA -- galaxy pairs which extend to lower masses will be able to test this possibility. 




\subsection{Ionizing radiation from local stars and AGN}\label{s:fesc}
    

Our result that the CGM can be largely neutral requires it to be shielded from local ionizing sources such as young stars and AGN. This follows from the expected ionization rate of a young stellar population if absorption by the galaxy is neglected:
\begin{equation}
 \Gamma_{\rm SF} = \frac{Q_{\rm SF}\sigma_{\rm HI}}{4 \pi r^2} \sim 1.7\cdot10^{-9} \frac{\rm SFR}{10\msun\yr^{-1}} \left(\frac{r}{10\kpc}\right)^{-2} \s^{-1} ~,
 \end{equation}
where $\sigma_{\rm HI}\approx10^{-17.2}\cm^{2}$ is the \hi\ opacity to ionizing radiation and $Q_{\rm SF}\sim 3.3 \cdot 10^{53} ({\rm SFR}/{\rm M}_\odot\yr^{-1}) \s^{-1}$ is the emission rate of ionizing photons. The numerical value of $Q_{\rm SF}$ is evaluated based on \cite{Sternberg03} for an upper IMF mass cutoff of $120\msun$, and is consistent with $Q_{\rm SF}$ in the stellar population model used in FIRE.
Similarly, for an AGN 
\begin{equation}
\Gamma_{\rm AGN} = \frac{Q_{\rm AGN}\sigma_{\rm HI}}{4 \pi r^2} \sim 1.9\cdot10^{-8} \frac{L_{\rm AGN}}{10^{46}\erg\s^{-1}} \left(\frac{r}{10\kpc}\right)^{-2} \s^{-1} ,
 \end{equation}
where we assumed an ionizing photon luminosity $Q_{\rm AGN}=L_{\rm AGN} / 72\ev$ where $L_{\rm AGN}$ is the bolometric luminosity \citep[e.g.,][]{Telfer02} and $\sigma_{\hi}=10^{-17.6}\cm^2$ for the hard AGN spectrum. 
At radii $r\approx10\kpc$ the values of $\Gamma_{\rm SF}$ and $\Gamma_{\rm AGN}$ are factors of $\sim10^3-10^4$ larger than $\Gamma_{\rm UVB}\sim10^{-12}\s^{-1}$
\changed{
while the extent of shielded region extends as $\Gamma^{-1/3}$ (eqn.~\ref{e:Rshielded}), so the CGM can be largely neutral only along directions in which local radiation is absorbed.  
}
\newcommand{\fesc}{f_{\rm esc}}


The FIRE simulations account for ionizing photons from young stars and for absorption of these photons by particles neighboring the star particle (section~\ref{s:fire}). In most cases where we find the inner CGM is neutral the neighboring particles entirely absorb the ionizing stellar radiation. This is evident from the extent of the neutral CGM being well approximated by $R(\nH=\nhss)$ (Fig.~\ref{f:fHI}), where $\nhss$ is calculated assuming ionization by local stars is negligible. 

\changed{
A more accurate calculation of the escape fraction $\fesc$ of ionizing photons from the galaxy was conducted by \cite{Ma20}, albeit in FIRE simulations with somewhat higher redshifts ($z\gtrsim5$) than those analyzed here. They post-processed the FIRE simulation snapshots with a Monte Carlo radiation transfer code, and found that the absorption of ionizing photons from young stars is bi-modal, with $\fesc$ approaching unity along low-column density paths from the most vigoursly forming star forming regions, while along other paths $\fesc\approx0$ (see their figure~8). It is thus conceivable that only a small fraction of the CGM is exposed to stellar ionizing radiation even if the average escape fraction is $\langle\fesc\rangle\sim10\%$ \citep[e.g.,][]{Steidel18}. To test this possibility we ran a preliminary Monte Carlo analysis on $z\sim4$ snapshots with $\Mhalo\sim10^{11.5}\msun$ in our sample, which have $\tcoolsh<\tff$ in their inner CGM (Fig.~\ref{f:fV cool}). We find the average $\fhi$ in the inner CGM are lower by $\lesssim25\%$ than in the fiducial calculation, suggesting our conclusions hold also when applying a more accurate radiation transfer calculation. A more thorough investigation is deferred to future work (Holguin et al., in prep.). 
}

In active galaxies the radiation from the central source is absorbed along $\sim50\%$ of sightlines,  giving rise to the type 1 AGN / type 2 AGN dichotomy, with the covering factor of the absorbing gas potentially increasing to lower AGN luminosities \citep{Maiolino07,Stern12b,Lusso13}. Thus, the CGM of active galaxies can be largely neutral only along type 2 directions. AGN are however plausibly rare enough not to affect our conclusions regarding typical CGM. Even relatively weak AGN with X-ray luminosities $\approx10^{42}\ergs$ have an upper limit on their number density of  $\sim10^{-4}\Mpc^{-3}$  at $z \sim 4$ \citep{Vito16}, 
\changed{
compared to a number density of $10^{-3}\Mpc^{-3}$ for $>10^{12}\msun$ halos and $0.03\Mpc^{-3}$ for $>10^{11}\msun$ halos \citep{RodriguezPuebla16}. 
}
%


\section{Summary}\label{s:summary}

We use the FIRE-2 cosmological simulations to study when the circumgalactic medium of galaxies is predominantly neutral and has a large DLA covering factor. 
We utilize a suite of zoom simulations which together span halo masses $10^{10}\lesssim\Mhalo\lesssim10^{13}\msun$ at $0<z<5$. Our results can be summarized as follows:

\begin{enumerate}
 \item The volume filling fraction of cool $T<10^{4.5}\K$ gas is large ($\fcool\gtrsim0.3$) at CGM radii where the cooling time of hot $T\sim\Tvir$ gas is shorter than the free-fall time (Fig.~\ref{f:fV cool}). This condition is satisfied in FIRE in the inner regions of halos with a mass of $\sim10^{12}\msun$ or lower, and extends to the outer regions of halos with a mass of $\sim10^{11}\msun$ or lower (Fig.~\ref{f:Rsonic}). 
 
 \item In $\lesssim10^{12}\msun$ FIRE halos where a large fraction of the inner CGM volume is occupied by cool gas, this gas is also neutral at radii where densities are large enough to shield it from photoionization. The maximum radius where the cool CGM is shielded increases relative to the halo size with redshift, roughly as $\sim0.3 ((1+z)/5)^{3/2}\Rvir$ (Fig.~\ref{f:fHI}).
 
 \changed{
 \item Gas radial velocities in predominantly neutral inner CGM in FIRE typically span $\pm\vvir$, with a small mean inflow velocity ($<0.2\vvir$, Fig.~\ref{f:vr}), suggesting a mixture of inflowing and outflowing gas. The importance of galaxy outflows for the neutral inner CGM is also indicated by its high metallicity, which is within a factor of two of the ISM metallicity in the central galaxy (Fig.~\ref{f:Z}).
}

 \item Predominantly neutral inner CGM have a DLA covering fraction approaching unity (Fig.~\ref{f:ratio}). This implies that at $z\sim4$ DLAs extend to radii that are an order of magnitude larger than the effective radius of the central galaxy. 
 \changed{
 Our conclusions on the extent of DLAs in the CGM are quantitatively similar to those of  \cite{ZhengMiraldaEscude02} and \cite{Theuns21}, who assumed a spherical CGM in photoionization equilibrium with the UVB. 
} 
 
 \item The FIRE simulations predict that $10^{10}-10^{11}\msun$ halos contribute a larger fraction of DLAs than more massive halos, but high-metallicity DLAs with $Z>0.1\zsun$ are dominated by massive halos with $\Mhalo>10^{11.5}\msun$ (Fig.~\ref{f:dNdXdlogZ}). 
 \changed{
 We also predict that the DLA covering fraction decreases at $\Mhalo\gg10^{12}\msun$ since the inner CGM of massive halos is filled with hot collisionally-ionized gas (Fig.~\ref{f:ratio}). 
 }
 
 \changed{
 \item Comparison of FIRE predictions with the DLA -- galaxy impact parameters recently observed by ALMA at $z\sim4$ \citep{Neeleman17,Neeleman19}, together with the observed DLA metallicities, velocity widths, and galaxy SFRs, suggests that these DLAs originate in a predominantly neutral inner CGM of $10^{11.5}-10^{12}\msun$ halos (Fig.~\ref{f:Z}).
 }
\end{enumerate}

If the high-redshift inner CGM is largely neutral, as we find in the FIRE simulations, then its mass, metallicity, and dust-to-gas ratio can be directly constrained by DLA surveys without resorting to large ionization corrections ($\sim10^3-10^4$) as required in the study of ionized low-$z$ CGM (e.g., \citealt{Werk14}). The high-redshift neutral CGM thus inherently avoids a major observational challenge of accounting for the hard-to-see ionized phase. 
As the CGM is the destination of galaxy-scale outflows, this possibility to accurately deduce its properties from high-$z$ observations may allow to better constrain the physics of galaxy feedback in the early Universe, a substantial uncertainty in current models of galaxy formation. 

\section*{Acknowledgements}

\changed{
We thank the anonymous referee for insightful comments which substantially improved the paper. We thank F.~Holguin and C.~Hayward for conducting the radiation transfer analysis described in section~\ref{s:fesc}. 
}
JS is supported by the CIERA Postdoctoral Fellowship Program. This research was also supported by the German Science Foundation via DIP grant STE~1869/2-1 GE~625/17-1 at Tel Aviv University. 
CAFG was supported by NSF through grants AST-1715216 and CAREER award AST-1652522; by NASA through grant 17-ATP17-0067; by STScI through grant HST-AR-16124.001-A; and by a Cottrell Scholar Award and a Scialog Award from the Research Corporation for Science Advancement.
DAA acknowledges support by NSF grant AST-2009687 and by the Flatiron Institute, which is supported by the Simons Foundation.
AW received support from NASA through ATP grants 80NSSC18K1097 and 80NSSC20K0513; HST grants GO-14734, AR-15057, AR-15809, and GO-15902 from STScI; a Scialog Award from the Heising-Simons Foundation; and a Hellman Fellowship.

\section*{Data availability}

The data supporting the plots within this article are available on reasonable request to the corresponding author. A public version of the GIZMO code is available at \url{http://www.tapir.caltech.edu/~phopkins/Site/GIZMO.html}.
Additional data including simulation snapshots, initial conditions, and derived data products are available at \url{https://fire.northwestern.edu/data/}.

\bibliographystyle{mnras}
\bibliography{main}

\begin{thebibliography}{}
\makeatletter
\relax
\def\mn@urlcharsother{\let\do\@makeother \do\$\do\&\do\#\do\^\do\_\do\%\do\~}
\def\mn@doi{\begingroup\mn@urlcharsother \@ifnextchar [ {\mn@doi@}
  {\mn@doi@[]}}
\def\mn@doi@[#1]#2{\def\@tempa{#1}\ifx\@tempa\@empty \href
  {http://dx.doi.org/#2} {doi:#2}\else \href {http://dx.doi.org/#2} {#1}\fi
  \endgroup}
\def\mn@eprint#1#2{\mn@eprint@#1:#2::\@nil}
\def\mn@eprint@arXiv#1{\href {http://arxiv.org/abs/#1} {{\tt arXiv:#1}}}
\def\mn@eprint@dblp#1{\href {http://dblp.uni-trier.de/rec/bibtex/#1.xml}
  {dblp:#1}}
\def\mn@eprint@#1:#2:#3:#4\@nil{\def\@tempa {#1}\def\@tempb {#2}\def\@tempc
  {#3}\ifx \@tempc \@empty \let \@tempc \@tempb \let \@tempb \@tempa \fi \ifx
  \@tempb \@empty \def\@tempb {arXiv}\fi \@ifundefined
  {mn@eprint@\@tempb}{\@tempb:\@tempc}{\expandafter \expandafter \csname
  mn@eprint@\@tempb\endcsname \expandafter{\@tempc}}}

\bibitem[\protect\citeauthoryear{{Angl{\'e}s-Alc{\'a}zar},
  {Faucher-Gigu{\`e}re}, {Kere{\v{s}}}, {Hopkins}, {Quataert}  \&
  {Murray}}{{Angl{\'e}s-Alc{\'a}zar} et~al.}{2017a}]{anglesalcazar17a}
{Angl{\'e}s-Alc{\'a}zar} D.,  {Faucher-Gigu{\`e}re} C.-A.,  {Kere{\v{s}}} D.,
  {Hopkins} P.~F.,  {Quataert} E.,   {Murray} N.,  2017a, \mn@doi [\mnras]
  {10.1093/mnras/stx1517}, \href
  {https://ui.adsabs.harvard.edu/abs/2017MNRAS.470.4698A} {470, 4698}

\bibitem[\protect\citeauthoryear{{Angl{\'e}s-Alc{\'a}zar},
  {Faucher-Gigu{\`e}re}, {Quataert}, {Hopkins}, {Feldmann}, {Torrey}, {Wetzel}
  \& {Kere{\v{s}}}}{{Angl{\'e}s-Alc{\'a}zar} et~al.}{2017b}]{anglesalcazar17b}
{Angl{\'e}s-Alc{\'a}zar} D.,  {Faucher-Gigu{\`e}re} C.-A.,  {Quataert} E.,
  {Hopkins} P.~F.,  {Feldmann} R.,  {Torrey} P.,  {Wetzel} A.,   {Kere{\v{s}}}
  D.,  2017b, \mn@doi [\mnras] {10.1093/mnrasl/slx161}, \href
  {https://ui.adsabs.harvard.edu/abs/2017MNRAS.472L.109A} {472, L109}

\bibitem[\protect\citeauthoryear{{Behroozi}, {Wechsler}, {Hearin}  \&
  {Conroy}}{{Behroozi} et~al.}{2019}]{Behroozi19}
{Behroozi} P.,  {Wechsler} R.~H.,  {Hearin} A.~P.,   {Conroy} C.,  2019,
  \mn@doi [\mnras] {10.1093/mnras/stz1182}, \href
  {https://ui.adsabs.harvard.edu/abs/2019MNRAS.488.3143B} {488, 3143}

\bibitem[\protect\citeauthoryear{{Berry}, {Somerville}, {Haas}, {Gawiser},
  {Maller}, {Popping}  \& {Trager}}{{Berry} et~al.}{2014}]{Berry14}
{Berry} M.,  {Somerville} R.~S.,  {Haas} M.~R.,  {Gawiser} E.,  {Maller} A.,
  {Popping} G.,   {Trager} S.~C.,  2014, \mn@doi [\mnras]
  {10.1093/mnras/stu613}, \href
  {https://ui.adsabs.harvard.edu/abs/2014MNRAS.441..939B} {441, 939}

\bibitem[\protect\citeauthoryear{{Bertschinger}}{{Bertschinger}}{1985}]{Bertschinger85}
{Bertschinger} E.,  1985, \mn@doi [\apjs] {10.1086/191028}, \href
  {https://ui.adsabs.harvard.edu/abs/1985ApJS...58...39B} {58, 39}

\bibitem[\protect\citeauthoryear{{Birnboim} \& {Dekel}}{{Birnboim} \&
  {Dekel}}{2003}]{Birnboim03}
{Birnboim} Y.,  {Dekel} A.,  2003, \mn@doi [\mnras]
  {10.1046/j.1365-8711.2003.06955.x}, \href
  {https://ui.adsabs.harvard.edu/abs/2003MNRAS.345..349B} {345, 349}

\bibitem[\protect\citeauthoryear{{Bryan} \& {Norman}}{{Bryan} \&
  {Norman}}{1998}]{bryan98}
{Bryan} G.~L.,  {Norman} M.~L.,  1998, \mn@doi [\apj] {10.1086/305262}, \href
  {https://ui.adsabs.harvard.edu/abs/1998ApJ...495...80B} {495, 80}

\bibitem[\protect\citeauthoryear{{Chan}, {Kere{\v{s}}}, {Wetzel}, {Hopkins},
  {Faucher-Gigu{\`e}re}, {El-Badry}, {Garrison-Kimmel}  \&
  {Boylan-Kolchin}}{{Chan} et~al.}{2018}]{Chan18}
{Chan} T.~K.,  {Kere{\v{s}}} D.,  {Wetzel} A.,  {Hopkins} P.~F.,
  {Faucher-Gigu{\`e}re} C.~A.,  {El-Badry} K.,  {Garrison-Kimmel} S.,
  {Boylan-Kolchin} M.,  2018, \mn@doi [\mnras] {10.1093/mnras/sty1153}, \href
  {https://ui.adsabs.harvard.edu/abs/2018MNRAS.478..906C} {478, 906}

\bibitem[\protect\citeauthoryear{{Chen} \& {Lanzetta}}{{Chen} \&
  {Lanzetta}}{2003}]{ChenLanzetta03}
{Chen} H.-W.,  {Lanzetta} K.~M.,  2003, \mn@doi [\apj] {10.1086/378635}, \href
  {https://ui.adsabs.harvard.edu/abs/2003ApJ...597..706C} {597, 706}

\bibitem[\protect\citeauthoryear{{Di Gioia}, {Cristiani}, {De Lucia}  \&
  {Xie}}{{Di Gioia} et~al.}{2020}]{DiGioia20}
{Di Gioia} S.,  {Cristiani} S.,  {De Lucia} G.,   {Xie} L.,  2020, \mn@doi
  [\mnras] {10.1093/mnras/staa2067}, \href
  {https://ui.adsabs.harvard.edu/abs/2020MNRAS.497.2469D} {497, 2469}

\bibitem[\protect\citeauthoryear{{Diemer}, {More}  \& {Kravtsov}}{{Diemer}
  et~al.}{2013}]{Diemer13}
{Diemer} B.,  {More} S.,   {Kravtsov} A.~V.,  2013, \mn@doi [\apj]
  {10.1088/0004-637X/766/1/25}, \href
  {https://ui.adsabs.harvard.edu/abs/2013ApJ...766...25D} {766, 25}

\bibitem[\protect\citeauthoryear{{El-Badry} et~al.,}{{El-Badry}
  et~al.}{2018}]{elbadry18a}
{El-Badry} K.,  et~al., 2018, \mn@doi [\mnras] {10.1093/mnras/stx2482}, \href
  {https://ui.adsabs.harvard.edu/abs/2018MNRAS.473.1930E} {473, 1930}

\bibitem[\protect\citeauthoryear{{Escala} et~al.,}{{Escala}
  et~al.}{2018}]{Escala18}
{Escala} I.,  et~al., 2018, \mn@doi [\mnras] {10.1093/mnras/stx2858}, \href
  {https://ui.adsabs.harvard.edu/abs/2018MNRAS.474.2194E} {474, 2194}

\bibitem[\protect\citeauthoryear{{Faerman}, {Sternberg}  \& {McKee}}{{Faerman}
  et~al.}{2020}]{Faerman20}
{Faerman} Y.,  {Sternberg} A.,   {McKee} C.~F.,  2020, \mn@doi [\apj]
  {10.3847/1538-4357/ab7ffc}, \href
  {https://ui.adsabs.harvard.edu/abs/2020ApJ...893...82F} {893, 82}

\bibitem[\protect\citeauthoryear{{Faucher-Gigu{\`e}re}}{{Faucher-Gigu{\`e}re}}{2020}]{FaucherGiguere20}
{Faucher-Gigu{\`e}re} C.-A.,  2020, \mn@doi [\mnras] {10.1093/mnras/staa302},
  \href {https://ui.adsabs.harvard.edu/abs/2020MNRAS.493.1614F} {493, 1614}

\bibitem[\protect\citeauthoryear{{Faucher-Gigu{\`e}re} \&
  {Kere{\v{s}}}}{{Faucher-Gigu{\`e}re} \&
  {Kere{\v{s}}}}{2011}]{FaucherGiguere11b}
{Faucher-Gigu{\`e}re} C.-A.,  {Kere{\v{s}}} D.,  2011, \mn@doi [\mnras]
  {10.1111/j.1745-3933.2011.01018.x}, \href
  {https://ui.adsabs.harvard.edu/abs/2011MNRAS.412L.118F} {412, L118}

\bibitem[\protect\citeauthoryear{{Faucher-Gigu{\`e}re}, {Lidz}, {Zaldarriaga}
  \& {Hernquist}}{{Faucher-Gigu{\`e}re} et~al.}{2009}]{FaucherGiguere09}
{Faucher-Gigu{\`e}re} C.-A.,  {Lidz} A.,  {Zaldarriaga} M.,   {Hernquist} L.,
  2009, \mn@doi [\apj] {10.1088/0004-637X/703/2/1416}, \href
  {https://ui.adsabs.harvard.edu/abs/2009ApJ...703.1416F} {703, 1416}

\bibitem[\protect\citeauthoryear{{Faucher-Gigu{\`e}re}, {Kere{\v{s}}},
  {Dijkstra}, {Hernquist}  \& {Zaldarriaga}}{{Faucher-Gigu{\`e}re}
  et~al.}{2010}]{FaucherGiguere10}
{Faucher-Gigu{\`e}re} C.-A.,  {Kere{\v{s}}} D.,  {Dijkstra} M.,  {Hernquist}
  L.,   {Zaldarriaga} M.,  2010, \mn@doi [\apj] {10.1088/0004-637X/725/1/633},
  \href {https://ui.adsabs.harvard.edu/abs/2010ApJ...725..633F} {725, 633}

\bibitem[\protect\citeauthoryear{{Faucher-Gigu{\`e}re}, {Hopkins},
  {Kere{\v{s}}}, {Muratov}, {Quataert}  \& {Murray}}{{Faucher-Gigu{\`e}re}
  et~al.}{2015}]{FaucherGiguere15}
{Faucher-Gigu{\`e}re} C.-A.,  {Hopkins} P.~F.,  {Kere{\v{s}}} D.,  {Muratov}
  A.~L.,  {Quataert} E.,   {Murray} N.,  2015, \mn@doi [\mnras]
  {10.1093/mnras/stv336}, \href
  {https://ui.adsabs.harvard.edu/abs/2015MNRAS.449..987F} {449, 987}

\bibitem[\protect\citeauthoryear{{Faucher-Gigu{\`e}re}, {Feldmann}, {Quataert},
  {Kere{\v{s}}}, {Hopkins}  \& {Murray}}{{Faucher-Gigu{\`e}re}
  et~al.}{2016}]{FaucherGiguere16}
{Faucher-Gigu{\`e}re} C.-A.,  {Feldmann} R.,  {Quataert} E.,  {Kere{\v{s}}} D.,
   {Hopkins} P.~F.,   {Murray} N.,  2016, \mn@doi [\mnras]
  {10.1093/mnrasl/slw091}, \href
  {https://ui.adsabs.harvard.edu/abs/2016MNRAS.461L..32F} {461, L32}

\bibitem[\protect\citeauthoryear{{Fielding}, {Quataert}, {McCourt}  \&
  {Thompson}}{{Fielding} et~al.}{2017}]{Fielding17}
{Fielding} D.,  {Quataert} E.,  {McCourt} M.,   {Thompson} T.~A.,  2017,
  \mn@doi [\mnras] {10.1093/mnras/stw3326}, \href
  {https://ui.adsabs.harvard.edu/abs/2017MNRAS.466.3810F} {466, 3810}

\bibitem[\protect\citeauthoryear{{Fumagalli}, {Prochaska}, {Kasen}, {Dekel},
  {Ceverino}  \& {Primack}}{{Fumagalli} et~al.}{2011}]{Fumagalli11}
{Fumagalli} M.,  {Prochaska} J.~X.,  {Kasen} D.,  {Dekel} A.,  {Ceverino} D.,
  {Primack} J.~R.,  2011, \mn@doi [\mnras] {10.1111/j.1365-2966.2011.19599.x},
  \href {https://ui.adsabs.harvard.edu/abs/2011MNRAS.418.1796F} {418, 1796}

\bibitem[\protect\citeauthoryear{{Fynbo}, {Prochaska}, {Sommer-Larsen},
  {Dessauges-Zavadsky}  \& {M{\o}ller}}{{Fynbo} et~al.}{2008}]{Fynbo08}
{Fynbo} J. P.~U.,  {Prochaska} J.~X.,  {Sommer-Larsen} J.,
  {Dessauges-Zavadsky} M.,   {M{\o}ller} P.,  2008, \mn@doi [\apj]
  {10.1086/589555}, \href
  {https://ui.adsabs.harvard.edu/abs/2008ApJ...683..321F} {683, 321}

\bibitem[\protect\citeauthoryear{{Garrison-Kimmel} et~al.,}{{Garrison-Kimmel}
  et~al.}{2017}]{GarrisonKimmel17}
{Garrison-Kimmel} S.,  et~al., 2017, \mn@doi [\mnras] {10.1093/mnras/stx1710},
  \href {https://ui.adsabs.harvard.edu/abs/2017MNRAS.471.1709G} {471, 1709}

\bibitem[\protect\citeauthoryear{{Garrison-Kimmel} et~al.,}{{Garrison-Kimmel}
  et~al.}{2018}]{GarrisonKimmel18}
{Garrison-Kimmel} S.,  et~al., 2018, \mn@doi [\mnras] {10.1093/mnras/sty2513},
  \href {https://ui.adsabs.harvard.edu/abs/2018MNRAS.481.4133G} {481, 4133}

\bibitem[\protect\citeauthoryear{{Haehnelt}, {Steinmetz}  \&
  {Rauch}}{{Haehnelt} et~al.}{1998}]{Haehnelt98}
{Haehnelt} M.~G.,  {Steinmetz} M.,   {Rauch} M.,  1998, \mn@doi [\apj]
  {10.1086/305323}, \href
  {https://ui.adsabs.harvard.edu/abs/1998ApJ...495..647H} {495, 647}

\bibitem[\protect\citeauthoryear{{Hafen} et~al.,}{{Hafen}
  et~al.}{2017}]{Hafen17}
{Hafen} Z.,  et~al., 2017, \mn@doi [\mnras] {10.1093/mnras/stx952}, \href
  {https://ui.adsabs.harvard.edu/abs/2017MNRAS.469.2292H} {469, 2292}

\bibitem[\protect\citeauthoryear{{Hafen} et~al.,}{{Hafen}
  et~al.}{2019}]{Hafen19}
{Hafen} Z.,  et~al., 2019, \mn@doi [\mnras] {10.1093/mnras/stz1773}, \href
  {https://ui.adsabs.harvard.edu/abs/2019MNRAS.488.1248H} {488, 1248}

\bibitem[\protect\citeauthoryear{{Hafen} et~al.,}{{Hafen}
  et~al.}{2020}]{Hafen20}
{Hafen} Z.,  et~al., 2020, \mn@doi [\mnras] {10.1093/mnras/staa902}, \href
  {https://ui.adsabs.harvard.edu/abs/2020MNRAS.494.3581H} {494, 3581}

\bibitem[\protect\citeauthoryear{{Ho}, {Bird}  \& {Garnett}}{{Ho}
  et~al.}{2020}]{Ho20}
{Ho} M.-F.,  {Bird} S.,   {Garnett} R.,  2020, \mn@doi [\mnras]
  {10.1093/mnras/staa1806}, \href
  {https://ui.adsabs.harvard.edu/abs/2020MNRAS.496.5436H} {496, 5436}

\bibitem[\protect\citeauthoryear{{Hopkins}}{{Hopkins}}{2015}]{Hopkins15}
{Hopkins} P.~F.,  2015, \mn@doi [\mnras] {10.1093/mnras/stv195}, \href
  {https://ui.adsabs.harvard.edu/abs/2015MNRAS.450...53H} {450, 53}

\bibitem[\protect\citeauthoryear{{Hopkins}}{{Hopkins}}{2017}]{Hopkins17}
{Hopkins} P.~F.,  2017, \mn@doi [\mnras] {10.1093/mnras/stw3306}, \href
  {https://ui.adsabs.harvard.edu/abs/2017MNRAS.466.3387H} {466, 3387}

\bibitem[\protect\citeauthoryear{{Hopkins}, {Kere{\v{s}}}, {O{\~n}orbe},
  {Faucher-Gigu{\`e}re}, {Quataert}, {Murray}  \& {Bullock}}{{Hopkins}
  et~al.}{2014}]{Hopkins14}
{Hopkins} P.~F.,  {Kere{\v{s}}} D.,  {O{\~n}orbe} J.,  {Faucher-Gigu{\`e}re}
  C.-A.,  {Quataert} E.,  {Murray} N.,   {Bullock} J.~S.,  2014, \mn@doi
  [\mnras] {10.1093/mnras/stu1738}, \href
  {https://ui.adsabs.harvard.edu/abs/2014MNRAS.445..581H} {445, 581}

\bibitem[\protect\citeauthoryear{{Hopkins} et~al.,}{{Hopkins}
  et~al.}{2018}]{Hopkins18}
{Hopkins} P.~F.,  et~al., 2018, \mn@doi [\mnras] {10.1093/mnras/sty1690}, \href
  {https://ui.adsabs.harvard.edu/abs/2018MNRAS.480..800H} {480, 800}

\bibitem[\protect\citeauthoryear{{Hummels}, {Smith}  \& {Silvia}}{{Hummels}
  et~al.}{2017}]{Hummels17}
{Hummels} C.~B.,  {Smith} B.~D.,   {Silvia} D.~W.,  2017, \mn@doi [\apj]
  {10.3847/1538-4357/aa7e2d}, \href
  {https://ui.adsabs.harvard.edu/abs/2017ApJ...847...59H} {847, 59}

\bibitem[\protect\citeauthoryear{{Klypin}, {Yepes}, {Gottl{\"o}ber}, {Prada}
  \& {He{\ss}}}{{Klypin} et~al.}{2016}]{Klypin16}
{Klypin} A.,  {Yepes} G.,  {Gottl{\"o}ber} S.,  {Prada} F.,   {He{\ss}} S.,
  2016, \mn@doi [\mnras] {10.1093/mnras/stw248}, \href
  {https://ui.adsabs.harvard.edu/abs/2016MNRAS.457.4340K} {457, 4340}

\bibitem[\protect\citeauthoryear{{Knollmann} \& {Knebe}}{{Knollmann} \&
  {Knebe}}{2009}]{Knollmann09}
{Knollmann} S.~R.,  {Knebe} A.,  2009, \mn@doi [\apjs]
  {10.1088/0067-0049/182/2/608}, \href
  {https://ui.adsabs.harvard.edu/abs/2009ApJS..182..608K} {182, 608}

\bibitem[\protect\citeauthoryear{{Kulkarni} \& {Fall}}{{Kulkarni} \&
  {Fall}}{2002}]{KulkarniFall02}
{Kulkarni} V.~P.,  {Fall} S.~M.,  2002, \mn@doi [\apj] {10.1086/343855}, \href
  {https://ui.adsabs.harvard.edu/abs/2002ApJ...580..732K} {580, 732}

\bibitem[\protect\citeauthoryear{{Ledoux}, {Petitjean}, {Fynbo}, {M{\o}ller}
  \& {Srianand}}{{Ledoux} et~al.}{2006}]{Ledoux06}
{Ledoux} C.,  {Petitjean} P.,  {Fynbo} J.~P.~U.,  {M{\o}ller} P.,   {Srianand}
  R.,  2006, \mn@doi [\aap] {10.1051/0004-6361:20054242}, \href
  {https://ui.adsabs.harvard.edu/abs/2006A&A...457...71L} {457, 71}

\bibitem[\protect\citeauthoryear{{Leitherer} et~al.,}{{Leitherer}
  et~al.}{1999}]{Leitherer99}
{Leitherer} C.,  et~al., 1999, \mn@doi [\apjs] {10.1086/313233}, \href
  {https://ui.adsabs.harvard.edu/abs/1999ApJS..123....3L} {123, 3}

\bibitem[\protect\citeauthoryear{{Lochhaas}, {Bryan}, {Li}, {Li}  \&
  {Fielding}}{{Lochhaas} et~al.}{2020}]{Lochhaas20}
{Lochhaas} C.,  {Bryan} G.~L.,  {Li} Y.,  {Li} M.,   {Fielding} D.,  2020,
  \mn@doi [\mnras] {10.1093/mnras/staa358}, \href
  {https://ui.adsabs.harvard.edu/abs/2020MNRAS.493.1461L} {493, 1461}

\bibitem[\protect\citeauthoryear{{Lusso} et~al.,}{{Lusso}
  et~al.}{2013}]{Lusso13}
{Lusso} E.,  et~al., 2013, \mn@doi [\apj] {10.1088/0004-637X/777/2/86}, \href
  {https://ui.adsabs.harvard.edu/abs/2013ApJ...777...86L} {777, 86}

\bibitem[\protect\citeauthoryear{{Ma} et~al.,}{{Ma} et~al.}{2018}]{ma18}
{Ma} X.,  et~al., 2018, \mn@doi [\mnras] {10.1093/mnras/sty1024}, \href
  {https://ui.adsabs.harvard.edu/abs/2018MNRAS.478.1694M} {478, 1694}

\bibitem[\protect\citeauthoryear{{Ma}, {Quataert}, {Wetzel}, {Hopkins},
  {Faucher-Gigu{\`e}re}  \& {Kere{\v{s}}}}{{Ma} et~al.}{2020}]{Ma20}
{Ma} X.,  {Quataert} E.,  {Wetzel} A.,  {Hopkins} P.~F.,  {Faucher-Gigu{\`e}re}
  C.-A.,   {Kere{\v{s}}} D.,  2020, \mn@doi [\mnras] {10.1093/mnras/staa2404},
  \href {https://ui.adsabs.harvard.edu/abs/2020MNRAS.498.2001M} {498, 2001}

\bibitem[\protect\citeauthoryear{{Maiolino}, {Shemmer}, {Imanishi}, {Netzer},
  {Oliva}, {Lutz}  \& {Sturm}}{{Maiolino} et~al.}{2007}]{Maiolino07}
{Maiolino} R.,  {Shemmer} O.,  {Imanishi} M.,  {Netzer} H.,  {Oliva} E.,
  {Lutz} D.,   {Sturm} E.,  2007, \mn@doi [\aap] {10.1051/0004-6361:20077252},
  \href {https://ui.adsabs.harvard.edu/abs/2007A&A...468..979M} {468, 979}

\bibitem[\protect\citeauthoryear{{Mathews} \& {Bregman}}{{Mathews} \&
  {Bregman}}{1978}]{Mathews78}
{Mathews} W.~G.,  {Bregman} J.~N.,  1978, \mn@doi [\apj] {10.1086/156379},
  \href {https://ui.adsabs.harvard.edu/abs/1978ApJ...224..308M} {224, 308}

\bibitem[\protect\citeauthoryear{{McCourt}, {Sharma}, {Quataert}  \&
  {Parrish}}{{McCourt} et~al.}{2012}]{McCourt12}
{McCourt} M.,  {Sharma} P.,  {Quataert} E.,   {Parrish} I.~J.,  2012, \mn@doi
  [\mnras] {10.1111/j.1365-2966.2011.19972.x}, \href
  {https://ui.adsabs.harvard.edu/abs/2012MNRAS.419.3319M} {419, 3319}

\bibitem[\protect\citeauthoryear{{Mo}, {Mao}  \& {White}}{{Mo}
  et~al.}{1998}]{Mo98}
{Mo} H.~J.,  {Mao} S.,   {White} S. D.~M.,  1998, \mn@doi [\mnras]
  {10.1046/j.1365-8711.1998.01227.x}, \href
  {https://ui.adsabs.harvard.edu/abs/1998MNRAS.295..319M} {295, 319}

\bibitem[\protect\citeauthoryear{{M{\o}ller} \& {Christensen}}{{M{\o}ller} \&
  {Christensen}}{2020}]{MollerChristensen20}
{M{\o}ller} P.,  {Christensen} L.,  2020, \mn@doi [\mnras]
  {10.1093/mnras/staa128}, \href
  {https://ui.adsabs.harvard.edu/abs/2020MNRAS.492.4805M} {492, 4805}

\bibitem[\protect\citeauthoryear{{Muratov}, {Kere{\v{s}}},
  {Faucher-Gigu{\`e}re}, {Hopkins}, {Quataert}  \& {Murray}}{{Muratov}
  et~al.}{2015}]{Muratov15}
{Muratov} A.~L.,  {Kere{\v{s}}} D.,  {Faucher-Gigu{\`e}re} C.-A.,  {Hopkins}
  P.~F.,  {Quataert} E.,   {Murray} N.,  2015, \mn@doi [\mnras]
  {10.1093/mnras/stv2126}, \href
  {https://ui.adsabs.harvard.edu/abs/2015MNRAS.454.2691M} {454, 2691}

\bibitem[\protect\citeauthoryear{{Neeleman}, {Kanekar}, {Prochaska},
  {Rafelski}, {Carilli}  \& {Wolfe}}{{Neeleman} et~al.}{2017}]{Neeleman17}
{Neeleman} M.,  {Kanekar} N.,  {Prochaska} J.~X.,  {Rafelski} M.,  {Carilli}
  C.~L.,   {Wolfe} A.~M.,  2017, \mn@doi [Science] {10.1126/science.aal1737},
  \href {https://ui.adsabs.harvard.edu/abs/2017Sci...355.1285N} {355, 1285}

\bibitem[\protect\citeauthoryear{{Neeleman}, {Kanekar}, {Prochaska}, {Rafelski}
   \& {Carilli}}{{Neeleman} et~al.}{2019}]{Neeleman19}
{Neeleman} M.,  {Kanekar} N.,  {Prochaska} J.~X.,  {Rafelski} M.~A.,
  {Carilli} C.~L.,  2019, \mn@doi [\apjl] {10.3847/2041-8213/aaf871}, \href
  {https://ui.adsabs.harvard.edu/abs/2019ApJ...870L..19N} {870, L19}

\bibitem[\protect\citeauthoryear{{Noterdaeme} et~al.,}{{Noterdaeme}
  et~al.}{2012}]{Noterdaeme12}
{Noterdaeme} P.,  et~al., 2012, \mn@doi [\aap] {10.1051/0004-6361/201220259},
  \href {https://ui.adsabs.harvard.edu/abs/2012A&A...547L...1N} {547, L1}

\bibitem[\protect\citeauthoryear{{Okamoto}, {Gao}  \& {Theuns}}{{Okamoto}
  et~al.}{2008}]{Okamoto08}
{Okamoto} T.,  {Gao} L.,   {Theuns} T.,  2008, \mn@doi [\mnras]
  {10.1111/j.1365-2966.2008.13830.x}, \href
  {https://ui.adsabs.harvard.edu/abs/2008MNRAS.390..920O} {390, 920}

\bibitem[\protect\citeauthoryear{{Parsotan}, {Cochrane}, {Hayward},
  {Angl{\'e}s-Alc{\'a}zar}, {Feldmann}, {Faucher-Gigu{\`e}re}, {Wellons}  \&
  {Hopkins}}{{Parsotan} et~al.}{2021}]{Parsotan21}
{Parsotan} T.,  {Cochrane} R.~K.,  {Hayward} C.~C.,  {Angl{\'e}s-Alc{\'a}zar}
  D.,  {Feldmann} R.,  {Faucher-Gigu{\`e}re} C.~A.,  {Wellons} S.,   {Hopkins}
  P.~F.,  2021, \mn@doi [\mnras] {10.1093/mnras/staa3765}, \href
  {https://ui.adsabs.harvard.edu/abs/2021MNRAS.501.1591P} {501, 1591}

\bibitem[\protect\citeauthoryear{{Planck Collaboration} et~al.,}{{Planck
  Collaboration} et~al.}{2018}]{planck18}
{Planck Collaboration} et~al., 2018, arXiv e-prints, \href
  {https://ui.adsabs.harvard.edu/abs/2018arXiv180706209P} {p. arXiv:1807.06209}

\bibitem[\protect\citeauthoryear{{Pontzen} et~al.,}{{Pontzen}
  et~al.}{2008}]{Pontzen08}
{Pontzen} A.,  et~al., 2008, \mn@doi [\mnras]
  {10.1111/j.1365-2966.2008.13782.x}, \href
  {https://ui.adsabs.harvard.edu/abs/2008MNRAS.390.1349P} {390, 1349}

\bibitem[\protect\citeauthoryear{{Prochaska} \& {Wolfe}}{{Prochaska} \&
  {Wolfe}}{1997}]{ProchaskaWolfe97}
{Prochaska} J.~X.,  {Wolfe} A.~M.,  1997, \mn@doi [\apj] {10.1086/304591},
  \href {https://ui.adsabs.harvard.edu/abs/1997ApJ...487...73P} {487, 73}

\bibitem[\protect\citeauthoryear{{Prochaska} \& {Wolfe}}{{Prochaska} \&
  {Wolfe}}{2009}]{ProchaskaWolfe09}
{Prochaska} J.~X.,  {Wolfe} A.~M.,  2009, \mn@doi [\apj]
  {10.1088/0004-637X/696/2/1543}, \href
  {https://ui.adsabs.harvard.edu/abs/2009ApJ...696.1543P} {696, 1543}

\bibitem[\protect\citeauthoryear{{Rafelski}, {Wolfe}, {Prochaska}, {Neeleman}
  \& {Mendez}}{{Rafelski} et~al.}{2012}]{Rafelski12}
{Rafelski} M.,  {Wolfe} A.~M.,  {Prochaska} J.~X.,  {Neeleman} M.,   {Mendez}
  A.~J.,  2012, \mn@doi [\apj] {10.1088/0004-637X/755/2/89}, \href
  {https://ui.adsabs.harvard.edu/abs/2012ApJ...755...89R} {755, 89}

\bibitem[\protect\citeauthoryear{{Rafelski}, {Neeleman}, {Fumagalli}, {Wolfe}
  \& {Prochaska}}{{Rafelski} et~al.}{2014}]{Rafelski14}
{Rafelski} M.,  {Neeleman} M.,  {Fumagalli} M.,  {Wolfe} A.~M.,   {Prochaska}
  J.~X.,  2014, \mn@doi [\apjl] {10.1088/2041-8205/782/2/L29}, \href
  {https://ui.adsabs.harvard.edu/abs/2014ApJ...782L..29R} {782, L29}

\bibitem[\protect\citeauthoryear{{Rahmani} et~al.,}{{Rahmani}
  et~al.}{2016}]{Rahmani16}
{Rahmani} H.,  et~al., 2016, \mn@doi [\mnras] {10.1093/mnras/stw1965}, \href
  {https://ui.adsabs.harvard.edu/abs/2016MNRAS.463..980R} {463, 980}

\bibitem[\protect\citeauthoryear{{Rahmati}, {Pawlik}, {Rai{\v{c}}evi{\'c}}  \&
  {Schaye}}{{Rahmati} et~al.}{2013}]{Rahmati13}
{Rahmati} A.,  {Pawlik} A.~H.,  {Rai{\v{c}}evi{\'c}} M.,   {Schaye} J.,  2013,
  \mn@doi [\mnras] {10.1093/mnras/stt066}, \href
  {https://ui.adsabs.harvard.edu/abs/2013MNRAS.430.2427R} {430, 2427}

\bibitem[\protect\citeauthoryear{{Rhodin}, {Agertz}, {Christensen}, {Renaud}
  \& {Fynbo}}{{Rhodin} et~al.}{2019}]{Rhodin19}
{Rhodin} N.~H.~P.,  {Agertz} O.,  {Christensen} L.,  {Renaud} F.,   {Fynbo}
  J.~P.~U.,  2019, \mn@doi [\mnras] {10.1093/mnras/stz1479}, \href
  {https://ui.adsabs.harvard.edu/abs/2019MNRAS.488.3634R} {488, 3634}

\bibitem[\protect\citeauthoryear{{Rodr{\'\i}guez-Puebla}, {Behroozi},
  {Primack}, {Klypin}, {Lee}  \& {Hellinger}}{{Rodr{\'\i}guez-Puebla}
  et~al.}{2016}]{RodriguezPuebla16}
{Rodr{\'\i}guez-Puebla} A.,  {Behroozi} P.,  {Primack} J.,  {Klypin} A.,  {Lee}
  C.,   {Hellinger} D.,  2016, \mn@doi [\mnras] {10.1093/mnras/stw1705}, \href
  {https://ui.adsabs.harvard.edu/abs/2016MNRAS.462..893R} {462, 893}

\bibitem[\protect\citeauthoryear{{Schaye}}{{Schaye}}{2001}]{Schaye01}
{Schaye} J.,  2001, \mn@doi [\apj] {10.1086/322421}, \href
  {https://ui.adsabs.harvard.edu/abs/2001ApJ...559..507S} {559, 507}

\bibitem[\protect\citeauthoryear{{Scott}}{{Scott}}{1992}]{Scott92}
{Scott} D.~W.,  1992, {Multivariate Density Estimation}

\bibitem[\protect\citeauthoryear{{Shibuya}, {Ouchi}  \& {Harikane}}{{Shibuya}
  et~al.}{2015}]{Shibuya15}
{Shibuya} T.,  {Ouchi} M.,   {Harikane} Y.,  2015, \mn@doi [\apjs]
  {10.1088/0067-0049/219/2/15}, \href
  {https://ui.adsabs.harvard.edu/abs/2015ApJS..219...15S} {219, 15}

\bibitem[\protect\citeauthoryear{{Springel}}{{Springel}}{2005}]{Springel05}
{Springel} V.,  2005, \mn@doi [\mnras] {10.1111/j.1365-2966.2005.09655.x},
  \href {https://ui.adsabs.harvard.edu/abs/2005MNRAS.364.1105S} {364, 1105}

\bibitem[\protect\citeauthoryear{{Steidel}, {Bogosavljevi{\'c}}, {Shapley},
  {Reddy}, {Rudie}, {Pettini}, {Trainor}  \& {Strom}}{{Steidel}
  et~al.}{2018}]{Steidel18}
{Steidel} C.~C.,  {Bogosavljevi{\'c}} M.,  {Shapley} A.~E.,  {Reddy} N.~A.,
  {Rudie} G.~C.,  {Pettini} M.,  {Trainor} R.~F.,   {Strom} A.~L.,  2018,
  \mn@doi [\apj] {10.3847/1538-4357/aaed28}, \href
  {https://ui.adsabs.harvard.edu/abs/2018ApJ...869..123S} {869, 123}

\bibitem[\protect\citeauthoryear{{Stern} \& {Laor}}{{Stern} \&
  {Laor}}{2012}]{Stern12b}
{Stern} J.,  {Laor} A.,  2012, \mn@doi [\mnras]
  {10.1111/j.1365-2966.2012.21772.x}, \href
  {https://ui.adsabs.harvard.edu/abs/2012MNRAS.426.2703S} {426, 2703}

\bibitem[\protect\citeauthoryear{{Stern}, {Fielding}, {Faucher-Gigu{\`e}re}  \&
  {Quataert}}{{Stern} et~al.}{2019}]{Stern19}
{Stern} J.,  {Fielding} D.,  {Faucher-Gigu{\`e}re} C.-A.,   {Quataert} E.,
  2019, \mn@doi [\mnras] {10.1093/mnras/stz1859}, \href
  {https://ui.adsabs.harvard.edu/abs/2019MNRAS.488.2549S} {488, 2549}

\bibitem[\protect\citeauthoryear{{Stern}, {Fielding}, {Faucher-Gigu{\`e}re}  \&
  {Quataert}}{{Stern} et~al.}{2020}]{Stern20a}
{Stern} J.,  {Fielding} D.,  {Faucher-Gigu{\`e}re} C.-A.,   {Quataert} E.,
  2020, \mn@doi [\mnras] {10.1093/mnras/staa198}, \href
  {https://ui.adsabs.harvard.edu/abs/2020MNRAS.492.6042S} {492, 6042}

\bibitem[\protect\citeauthoryear{{Stern} et~al.,}{{Stern}
  et~al.}{2021}]{Stern20b}
{Stern} J.,  et~al., 2021, \mn@doi [\apj] {10.3847/1538-4357/abd776}, \href
  {https://ui.adsabs.harvard.edu/abs/2021ApJ...911...88S} {911, 88}

\bibitem[\protect\citeauthoryear{{Sternberg}, {Hoffmann}  \&
  {Pauldrach}}{{Sternberg} et~al.}{2003}]{Sternberg03}
{Sternberg} A.,  {Hoffmann} T.~L.,   {Pauldrach} A.~W.~A.,  2003, \mn@doi
  [\apj] {10.1086/379506}, \href
  {https://ui.adsabs.harvard.edu/abs/2003ApJ...599.1333S} {599, 1333}

\bibitem[\protect\citeauthoryear{{Telfer}, {Zheng}, {Kriss}  \&
  {Davidsen}}{{Telfer} et~al.}{2002}]{Telfer02}
{Telfer} R.~C.,  {Zheng} W.,  {Kriss} G.~A.,   {Davidsen} A.~F.,  2002, \mn@doi
  [\apj] {10.1086/324689}, \href
  {https://ui.adsabs.harvard.edu/abs/2002ApJ...565..773T} {565, 773}

\bibitem[\protect\citeauthoryear{{Theuns}}{{Theuns}}{2021}]{Theuns21}
{Theuns} T.,  2021, \mn@doi [\mnras] {10.1093/mnras/staa3412}, \href
  {https://ui.adsabs.harvard.edu/abs/2021MNRAS.500.2741T} {500, 2741}

\bibitem[\protect\citeauthoryear{{Tumlinson}, {Peeples}  \& {Werk}}{{Tumlinson}
  et~al.}{2017}]{Tumlinson17}
{Tumlinson} J.,  {Peeples} M.~S.,   {Werk} J.~K.,  2017, \mn@doi [\araa]
  {10.1146/annurev-astro-091916-055240}, \href
  {https://ui.adsabs.harvard.edu/abs/2017ARA&A..55..389T} {55, 389}

\bibitem[\protect\citeauthoryear{{Vito} et~al.,}{{Vito} et~al.}{2016}]{Vito16}
{Vito} F.,  et~al., 2016, \mn@doi [\mnras] {10.1093/mnras/stw1998}, \href
  {https://ui.adsabs.harvard.edu/abs/2016MNRAS.463..348V} {463, 348}

\bibitem[\protect\citeauthoryear{{Wellons}, {Faucher-Gigu{\`e}re},
  {Angl{\'e}s-Alc{\'a}zar}, {Hayward}, {Feldmann}, {Hopkins}  \&
  {Kere{\v{s}}}}{{Wellons} et~al.}{2020}]{Wellons20}
{Wellons} S.,  {Faucher-Gigu{\`e}re} C.-A.,  {Angl{\'e}s-Alc{\'a}zar} D.,
  {Hayward} C.~C.,  {Feldmann} R.,  {Hopkins} P.~F.,   {Kere{\v{s}}} D.,  2020,
  \mn@doi [\mnras] {10.1093/mnras/staa2229}, \href
  {https://ui.adsabs.harvard.edu/abs/2020MNRAS.497.4051W} {497, 4051}

\bibitem[\protect\citeauthoryear{{Werk} et~al.,}{{Werk} et~al.}{2014}]{Werk14}
{Werk} J.~K.,  et~al., 2014, \mn@doi [\apj] {10.1088/0004-637X/792/1/8}, \href
  {https://ui.adsabs.harvard.edu/abs/2014ApJ...792....8W} {792, 8}

\bibitem[\protect\citeauthoryear{{Wetzel} \& {Nagai}}{{Wetzel} \&
  {Nagai}}{2015}]{WetzelNagai15}
{Wetzel} A.~R.,  {Nagai} D.,  2015, \mn@doi [\apj]
  {10.1088/0004-637X/808/1/40}, \href
  {https://ui.adsabs.harvard.edu/abs/2015ApJ...808...40W} {808, 40}

\bibitem[\protect\citeauthoryear{{Wetzel}, {Hopkins}, {Kim},
  {Faucher-Gigu{\`e}re}, {Kere{\v{s}}}  \& {Quataert}}{{Wetzel}
  et~al.}{2016}]{Wetzel16}
{Wetzel} A.~R.,  {Hopkins} P.~F.,  {Kim} J.-h.,  {Faucher-Gigu{\`e}re} C.-A.,
  {Kere{\v{s}}} D.,   {Quataert} E.,  2016, \mn@doi [\apjl]
  {10.3847/2041-8205/827/2/L23}, \href
  {https://ui.adsabs.harvard.edu/abs/2016ApJ...827L..23W} {827, L23}

\bibitem[\protect\citeauthoryear{{Wiersma}, {Schaye}  \& {Smith}}{{Wiersma}
  et~al.}{2009}]{Wiersma09}
{Wiersma} R. P.~C.,  {Schaye} J.,   {Smith} B.~D.,  2009, \mn@doi [\mnras]
  {10.1111/j.1365-2966.2008.14191.x}, \href
  {https://ui.adsabs.harvard.edu/abs/2009MNRAS.393...99W} {393, 99}

\bibitem[\protect\citeauthoryear{{Wolfe}, {Gawiser}  \& {Prochaska}}{{Wolfe}
  et~al.}{2005}]{Wolfe05}
{Wolfe} A.~M.,  {Gawiser} E.,   {Prochaska} J.~X.,  2005, \mn@doi [\araa]
  {10.1146/annurev.astro.42.053102.133950}, \href
  {https://ui.adsabs.harvard.edu/abs/2005ARA&A..43..861W} {43, 861}

\bibitem[\protect\citeauthoryear{{Zheng} \& {Miralda-Escud{\'e}}}{{Zheng} \&
  {Miralda-Escud{\'e}}}{2002}]{ZhengMiraldaEscude02}
{Zheng} Z.,  {Miralda-Escud{\'e}} J.,  2002, \mn@doi [\apjl] {10.1086/340330},
  \href {https://ui.adsabs.harvard.edu/abs/2002ApJ...568L..71Z} {568, L71}

\makeatother
\end{thebibliography}

\appendix

\section{Jeans Scale approximation}\label{a:Jeans}

The Jeans scale approximation for $\nhss$ (eqn.~\ref{e:nhss}), is based on the assumption that absorbers are in hydrostatic equilibrium and have a gas fraction equal to the cosmic baryon fraction $f_{\rm g}=\Omega_{\rm b}/\Omega_{\rm m}\approx0.16$ \citep{Schaye01}. We repeat Schaye's derivation here for reference. 

Writing hydrostatic equilibrium as an equality between the dynamical time and the sound-crossing time we get
\begin{equation}
 \sqrt{\frac{f_{\rm g}}{G\rho}} = L \sqrt{\frac{\mu m_{\rm p}}{\gamma kT}} ~,
\end{equation}
where $L=\NH/\nH$ is the absorber size and $\gamma=5/3$ is the adiabatic index. Using this relation and the hydrogen mass fraction $X\equiv \nH m_{\rm p}/\rho$ we can derive the Jeans column density
\begin{eqnarray}
 N_{\rm H;\ J} &=& \left(\frac{\gamma k X f_{\rm g}}{\mu m_{\rm p}^2 G}\right)^{1/2}\nH^{1/2} T^{1/2} \nonumber \\ &\sim&  1.5\times 10^{20} T_4^{1/2} \left(\frac{\nH}{0.01\cm^{-2}}\right)^{1/2}\left(\frac{f_{\rm g}}{0.16}\right)^{1/2} \cm^{-2} ~.\nonumber\\
\end{eqnarray}
For highly ionized optically thin gas we can approximate $\fhi=0.46 \,T_4^{-0.76}\,\Gamma_{-12}^{-1}(\nH/{\rm cm}^{-3})$, so from 
$N_{\rm HI;\ J} =\fhi N_{\rm H;\ J}$ we get
\begin{equation}
 N_{\rm HI;\ J} \sim  6.9\times 10^{17} \,T_4^{-0.26}\Gamma_{-12}^{-1}\left(\frac{\nH}{0.01\cm^{-2}}\right)^{3/2}\left(\frac{f_{\rm g}}{0.16}\right)^{1/2} \cm^{-2}~.
\end{equation}
The shielding density at which the cloud becomes optical thick can then be derived from requiring $\sigma_\hi N_{\rm HI;\ J}=1$, where $\sigma_\hi$ is the opacity of \hi\ to ionizing photons. This gives
\begin{equation}
 \nhss \sim 0.01 \times T_4^{0.173}\Gamma_{-12}^{2/3}\left(\frac{f_{\rm g}}{0.16}\right)^{-1/3} \left(\frac{\sigma_\hi}{10^{-17.6}\cm^2}\right)^{-2/3}\cm^{-3} ~.
\end{equation}

\section{Analytic estimates of CGM densities and columns}\label{a:NH}

Characteristic CGM columns and densities can be derived by assuming a spherically-symmetric CGM with a power-law density profile $-a$:
\begin{equation}\label{e:density profile}
 \rho=\rho(\Rvir)\left(\frac{r}{\Rvir}\right)^{-a}  ~.
\end{equation}
The normalization of the density profile is chosen so that the total gas mass within $\Rvir$ equals $M_{\rm gas}=\fg(\Omega_{\rm b}/\Omega_{\rm m})\Mhalo$, where $\fg$ is the CGM mass relative to the halo cosmic baryon budget. Typical $\fg$ in FIRE, estimated by summing the gas mass at $0.1-1\Rvir$, are $\sim0.15-0.35$ at $z\sim0$ and $\approx0.4-0.8$ at $2<z<5$ (see also \citealt{Hafen19}). We thus get 
\begin{equation}\label{e:density normalization}
 \rho(\Rvir) = \left(1-\frac{a}{3}\right)\frac{\Omega_{\rm b}}{\Omega_{\rm m}}\fg\Deltac\rhocrit~,
\end{equation}
where $\rhocrit=3H^2/8\pi G$ is the critical density ($H$ is the Hubble parameter), and $\Deltac$ is the mean halo mass overdensity with respect to $\rhocrit$ \citep{bryan98}. Assuming a hydrogen density $\nH=0.7\rho/m_{\rm p}$ and $a=-2$ similar to the slopes in FIRE (Fig.~\ref{f:density}) we get for the inner CGM:
\begin{equation}\label{e:nH}
 \nH (r) = 0.13 \fg \left(\frac{1+z}{5}\right)^{3}   \left(\frac{r}{0.1\Rvir}\right)^{-2} \cm^{-3}  ~,
\end{equation}
where we used an Einstein-deSitter approximaton for $\sqrt{\Deltac} H$, which is accurate to $10\%$ or less at $z>1$ in our assumed cosmology:
\begin{equation}\label{e:Delta c H}
 \sqrt{\Deltac} H \approx 5.8 \left(\frac{1+z}{5}\right)^{3/2} \Gyr^{-1} ~.
\end{equation}
Using eqn.~(\ref{e:nH}) and the density threshold for shielding (eqn.~\ref{e:nhss}), one can estimate the shielding radius $\Rthick$ within which cool gas is shielded from ambient radiation. This gives:
\begin{equation}\label{e:Rshielded appendix}
 R(\nH=\nhss) = 0.33 \fg^{1/2}\left(\frac{1+z}{5}\right)^{3/2}\Gamma_{-12}^{-1/3}  T_4^{-0.09} \Rvir ~.
\end{equation}
In units of $\Rvir$, this radius has a strong dependence on redshift and no dependence on halo mass, similar to the comparison of mean densities with $\nhss$ in FIRE (Fig.~\ref{f:density}). 

Eqn.~(\ref{e:nH}) can be used to estimate the hydrogen column at an impact parameter $\Rimp$: 
\begin{eqnarray}\label{e:NH appendix}
 \NH(\Rimp) &=& \int \nH ds \nonumber \\ 
 &=& 8.3\cdot10^{21} \fg M_{12}^{1/3}\left(\frac{1+z}{5}\right)^{2}  \left(\frac{\Rimp}{0.1\Rvir}\right)^{-1} \cm^{-2} ~,\nonumber\\
\end{eqnarray}
where $ds=dr/\sqrt{1-(\Rimp/r)^2}$ and we calculated the integral using the equality $\int_1^\infty x^{-1}(x^2-1)^{-0.5}d x=\pi/2$. 
We also used the following approximation for the virial radius based on eqn.~(\ref{e:Delta c H}):
\begin{equation}\label{e:rvir}
 \Rvir = \left(\frac{2G\Mhalo}{\Deltac H^2}\right)^{1/3} \approx 64\, M_{12}^{1/3} \left(\frac{1+z}{5}\right)^{-1} \kpc ~.
\end{equation}
Alternative \changed{density profiles with} power-law slopes in the range $-2.8<a<-1.2$ yield columns which are within $50\%$ of the value in eqn.~(\ref{e:NH appendix}). 
Eqn.~(\ref{e:NH appendix}) demonstrates that for $\fg\sim0.5$ and halo masses above $\sim10^{9}\msun$, characteristic columns are substantially larger than the \hi-column threshold for DLAs of $2\cdot10^{20}\cm^{-2}$. A large neutral fraction is hence a sufficient condition for a large DLA covering factor in these halos. 
At halo masses $<10^{9}\msun$ where $\Tvir\lesssim10^4\K$ (see eqn.~\ref{e:Tvir}) halos are expected to lose their gas due to photoheating by the UVB \citep{Okamoto08}, and thus the volume-filling phase of their CGM is unlikely to significantly contribute to the DLA population.


We show also that our derivation of $R(\nH=\nhss)$ in FIRE is comparable to the idealized radiation transfer calculation of $\Rthick$ by ZME02. Defining $\RthickZME$ as the radius where the neutral hydrogen fraction in the ZME02 approaches unity (equal to $0.5 r_{\rm ss}$ in their notation, see their figure~1) we get $\RthickZME=14\kpc$ for their fiducial parameters of $\Mhalo=10^{12}\msun$, $M_{\rm gas}(<\Rvir)=0.05\Mhalo$, $\Rvir=51.5\kpc$, and $\Gamma_{-12}=1.25$.
Using their scaling of $\RthickZME\propto (f_{\rm gas}\Mvir/\rvir)^{2/3}\Gamma^{-1/3}$ and eqn.~(\ref{e:rvir}) this yields 
\begin{equation}
 \RthickZME = 0.43 \, M_{12}^{1/9}\left(\frac{1+z}{5}\right)^{3/2} \fg^{2/3}\Gamma_{-12}^{-1/3}  \Rvir  ~,
\end{equation}
which is comparable to eqn.~(\ref{e:Rshielded appendix}), and to the radius where $\fhi=0.5$ in FIRE (Fig.~\ref{f:fHI}).
This similarity is despite that the calculation of $\nhss$ is based on a Jeans-scale approximation for the size of the cool gas clouds, while the ZME02 calculation is based on radiation transfer in a spherically-symmetric cool CGM.

%

\section{Different physical implementations and resolution}

\begin{figure*}
\includegraphics{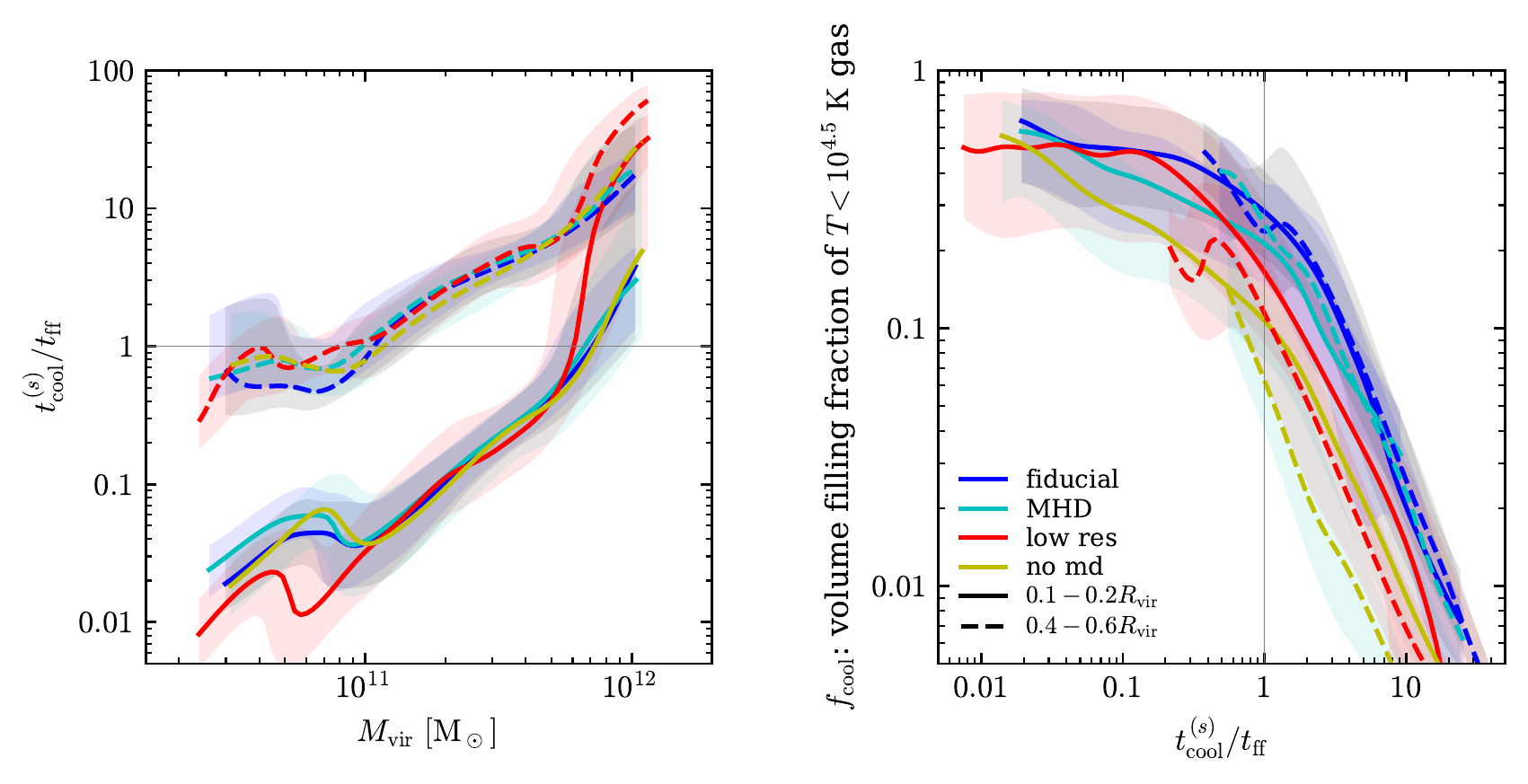}
\caption{
Similar to Fig.~\ref{f:fV cool} for different implementations of the m12i simulation. Blue lines denote the fiducial calculation used in the main text, which includes a prescription for subgrid metal diffusion and a gas mass resolution of $7100\msun$. Cyan lines denote a lower resolution calculation with gas mass resolution of $57000\msun$, red line denotes a MHD calculation, while the yellow line denotes a calculation that does not include subgrid metal diffusion. The left panel shows that the relation between $\tcoolsh/\tff$, halo mass and radius is rather independent of the inclusion of MHD, metal diffusion, or resolution at the tested range. The right panel shows that the relation between $\fcool$ and $\tcoolsh/\tff$ has a similar shape in all calculations, with the no metal-diffusion calculation offset to a factor of $2-3$ lower $\fcool$ at a given $\tcoolsh/\tff$. 
}
\label{f:fV cool appendix}
\end{figure*}

In Figure~\ref{f:fV cool appendix} we repeat the calculation of Fig.~\ref{f:fV cool} for different implementations of the m12i simulation, which has $\Mvir(z=0)\approx10^{12}\msun$. Blue lines denote the fiducial calculation used in the main text, which includes a prescription for subgrid metal diffusion \citep{Hopkins17,Escala18} and a gas mass resolution of $7100\msun$. Cyan lines denote a lower resolution calculation with gas mass of $57000\msun$, red lines denote an MHD calculation,
while the yellow line denotes a calculation which does not include subgrid metal diffusion. The left panel shows that the relation between $\tcoolsh/\tff$, halo mass and radius is rather independent of physical implementation, except in the low resolution simulation where $\tcoolsh/\tff$ increases above unity at somewhat lower halo masses than in the other simulations (see also fig.~21 in \citetalias{Stern20b}). The right panel shows that the shape of the relation between $\fcool$ and $\tcoolsh/\tff$ is similar across all calculations. The no metal-diffusion simulation is offset to a factor of $2-3$ lower $\fcool$ at a given $\tcoolsh/\tff$ compared to the other calculations, potentially since without subgrid diffusion metals and cooling are concentrated in a small fraction of CGM resolution elements \citep{Hafen19}. 

\label{lastpage}
\end{document}